\documentclass{jfm}
\usepackage{amsmath}
\usepackage{amsfonts}
\usepackage{bm}
\usepackage{enumerate}
\usepackage{graphicx}
\usepackage{mathrsfs}
\usepackage{subfigure}
\usepackage{url}
\usepackage{verbatim}
\usepackage{amssymb}
\usepackage[pdftex,bookmarksnumbered]{hyperref}
\usepackage{color}

\usepackage{epstopdf, epsfig}

\shorttitle{Diffusion-driven flows in a non-linear stratified fluid layer }

\title{Diffusion-driven flows in a non-linear stratified fluid layer}

\author{Lingyun Ding\aff{1}
  \corresp{\email{dingly@g.ucla.edu}}}

  \affiliation{\aff{1}Department of Mathematics, University of California Los Angeles, CA, 90095, United States}

\begin{document}
\maketitle

\begin{abstract}
  Diffusion-driven flow is a boundary layer flow arising from the interplay of gravity and diffusion in density-stratified fluids when a gravitational field is non-parallel to an impermeable solid boundary. This study investigates diffusion-driven flow within a nonlinearly density-stratified fluid confined between two tilted parallel walls. We introduce an asymptotic expansion inspired by the center manifold theory, where quantities are expanded in terms of derivatives of the cross-sectional averaged stratified scalar (such as salinity or temperature). This technique provides accurate approximations for velocity, density, and pressure fields. Furthermore, we derive an evolution equation describing the cross-sectional averaged stratified scalar. This equation takes the form of the traditional diffusion equation but replaces the constant diffusion coefficient with a positive-definite function dependent on the solution's derivative. Numerical simulations validate the accuracy of our approximations. Our investigation of the effective equation reveals that the density profile depends on a non-dimensional parameter denoted as $\gamma$ representing the flow strength. In the large $\gamma$ limit, the system is approximated by a diffusion process with an augmented diffusion coefficient of $1+\cot^{2}\theta$, where $\theta$ signifies the inclination angle of the channel domain. This parameter regime is where diffusion-driven flow exhibits its strongest mixing ability. Conversely, in the small $\gamma$ regime, the density field behaves like pure diffusion with distorted isopycnals. Lastly, we show that the classical thin film equation aligns with the results obtained using the proposed expansion in the small $\gamma$ regime but fails to accurately describe the dynamics of the density field for large $\gamma$.
\end{abstract}


\section{Introduction }
The density of a fluid is influenced by several factors, including temperature, the concentration of solute, and pressure profiles. Typically, these factors lead to a non-uniform distribution of density in the fluid. Density stratified fluids are commonly found in various natural environments, such as lakes, oceans, the Earth's atmosphere, making them a subject of great interest in numerous research endeavors  (\cite{linden1979mixing,cenedese2008mixing,magnaudet2020particles,camassa2022critical,more2023motion}). In the context of density-stratified fluids, achieving hydrostatic equilibrium is contingent upon aligning the density gradients with the direction of gravitational force.  In scenarios involving impermeable boundaries and the modeling of diffusive stratified  scalars,  a no-flux boundary condition applies, demanding that the density gradient be orthogonal to the boundary's normal vector. If the boundary's normal vector is perpendicular to the gravitational direction, the no-flux condition prevents the density gradient from aligning with gravity, thus disrupting hydrostatic equilibrium.  The disruption of hydrostatic equilibrium results in the emergence of a boundary layer flow, a phenomenon termed diffusion-driven flows in the field of physics (\cite{phillips1970flows, wunsch1970oceanic}), or mountain and valley (katabatic or anabatic) winds (\cite{prandtl1942fuhrer,oerlemans2002glacier}) in meteorology.

The phenomenon of diffusion-driven flow has garnered significant attention across various fields. Firstly, in the context of oceanography, the presence of salt in seawater leads to density stratification. The continental shelf has a gentle slope extending from the coast to the deeper waters of the ocean. The sloping boundaries induce a mean upwelling velocity along the boundaries to satisfy the no-flux condition. This upwelling flow plays a crucial role in facilitating the vertical exchange of oceanic properties (see, for example, \cite{phillips1970flows,wunsch1970oceanic,dell2015diffusive,drake2020abyssal,holmes2019tracer}). Secondly,  fluids confined within long, narrow fissures can also exhibit density stratification. This stratification can result from non-uniform concentration distributions of stratified  scalars or vertical temperature gradients. For example, Earth's geothermal gradient induces convective flows that lead to significant solute dispersion on geological timescales (\cite{woods1992natural,shaughnessy1995low,heitz2005optimizing}). Thirdly, when a wedge-shaped object is immersed in a stratified fluid, the resulting diffusion-driven flow can propel the object forward (\cite{mercier2014self,allshouse2010propulsion}). The flows generated by such immersed wedge-shaped objects have been further investigated in various contexts (\cite{chashechkin2004visual,zagumennyi2016diffusion,levitsky2019visualization,dimitrieva2019stratified,chashechkin2018waves}).  Fourthly, diffusion-driven flow is one of the mechanisms that induce particle attraction and self-assembly in stratified fluids (\cite{camassa2019first,thomas2023self}). Fifthly, it has been demonstrated that this type of flow can be employed to measure the molecular diffusivity of stratified  scalars (\cite{allshouse2010novel}). Sixthly,  flows induced by the presence of insulating sloping boundaries play an important role in the layer formation in double-diffusive systems (\cite{linden1977formation}).

Despite the significant findings in this field, there are two points that have not been adequately addressed in the existing literature. Firstly, most existing theories primarily pertain to linearly stratified fluids, and there is a scarcity of theoretical investigations into diffusion-driven flow in nonlinearly stratified fluids. Nonlinearly stratified fluids are more prevalent in various natural environments, making theoretical analysis in this context highly applicable. Secondly,  there can exist two different type of scalars in the fluid: the stratified  scalar and the passive scalar. The stratified  scalar causes non-uniform density distributions in the fluid and drives diffusion-driven flow. In contrast, the passive scalar represents the concentration of a different solute (in some cases, the temperature field) that doesn't contribute to density variation but is instead passively advected by the fluid flow.  It is well-established that fluid flow enhances solute mixing in the fluid (\cite{lin2011optimal,thiffeault2012using,aref2017frontiers}).  Many studies focus on how diffusion-driven flow enhances the dispersion of a passive scalar (\cite{woods1992natural,ding2023dispersion}) and the corresponding analysis for the stratified  scalar are rare.  Experiments, such as the one documented in (\cite{ding2022scalar}), illustrate that the dispersion of the stratified  scalar in a tilting capillary pipe exhibits a higher dispersion rate compared to a vertically oriented pipe. This qualitative demonstration highlights the role of diffusion-driven enhancement at micro-scales. At diffusion timescales, the concentration of a diffusing passive scalar under the advection of shear flows can be described by an effective diffusion equation with an effective diffusion coefficient (\cite{taylor1953dispersion,aris1956dispersion}). Consequently, it is of interest to establish the evolution equation for the concentration of the stratified  scalar. This would enable us to quantitatively describe how diffusion-driven flow enhances the spreading rate of the stratified  scalar.

To address these research gaps, this paper delves into the study of an incompressible viscous density stratified fluid layer confined between two infinitely parallel walls. Investigating this fundamental domain geometry can enhance our comprehension of more intricate shapes found in real-world scenarios, such as rock fissures and capillary pipes. Furthermore, it provides a versatile framework for comprehending fluid dynamics within confined spaces, which holds substantial practical significance.

Our primary objectives are to derive approximations for the velocity field and density field and to determine the equation governing the dynamics of the fluid density field at diffusion time scales. One conventional approach, such as thin film approximation, involves introducing a small parameter and utilizing power series expansions for all relevant quantities with respect to this parameter. However, we demonstrate that results obtained through thin film approximation are only valid within specific parameter ranges and do not provide uniformly accurate approximations across all parameters. To obtain an  approximation that accurately describe density dynamics across a wider parameter range, we adopt an alternative expansion method inspired by center manifold theory (\cite{mercer1990centre,roberts1996low,ding2022determinism}).  In this alternative approach, we assume that the density field varies slowly in the longitudinal direction of the channel. Then all quantities are represented in terms of derivatives of the averaged stratified  scalar. This innovative approach enhances our ability to achieve accuracy in a broader range of flow scenarios.

This paper is organized as follows: In Section \ref{sec:diffusion induce flow Governing equation and nondimensionalization}, we formulate the governing equations for diffusion-driven flow and outline the procedure for non-dimensionalization. Section \ref{sec:Flow for slowly varying density profile}  introduces an asymptotic expansion technique for the velocity, density, and pressure fields.  Using this  approach, we derive the leading-order approximation for the velocity field and establish the evolution equation for the cross-sectional averaged stratified  scalar, referred to as the effective equation in subsequent sections. In Section \ref{sec:Numerical simulation}, we conduct numerical simulations of the full governing equations to demonstrate the validity of the asymptotic approximations. Section \ref{sec:limit cases} delves deeper into the properties of the effective equation, providing a more comprehensive understanding of its behavior in various scenarios. Section \ref{sec:power series expansion} documents the results obtained using the thin film approximation and compares them with the expansion proposed in Section \ref{sec:Flow for slowly varying density profile}. Finally, in Section \ref{sec:conclusion}, we summarize our findings and explore potential avenues for future research.

\section{Governing equation and nondimensionalization }
\label{sec:diffusion induce flow Governing equation and nondimensionalization}
This section summarizes the mathematical formulation of the problem and documents the nondimensionalization procedure for the governing equation. 

\subsection{Governing equation}

\begin{figure}
  \centering
 \includegraphics[width=1\linewidth]{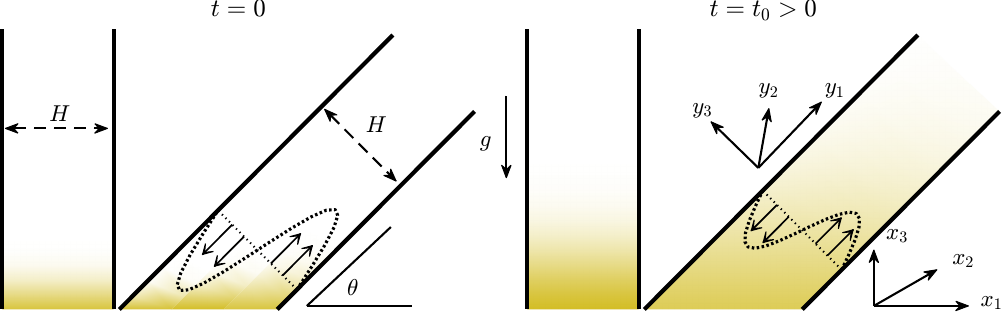}  
    \hfill
    \caption[]  { This schematic illustrates the setup for the diffusion-driven flow problem.  Tilted impermeable boundaries induce the diffusion driven flow, while there is no flow in the vertically oriented channel  ($\theta=\frac{\pi}{2}$). Pure molecular diffusion governs the scalar distribution in the latter case. As depicted in the left panel, at $t=0$, the concentration profiles are approximately the same in both channels. However, as time progresses, the dispersion of the scalar becomes more pronounced in the tilted channel compared to the vertically oriented one. In addition, the flow strength decreases as the density gradient decreases.    }
  \label{fig:DiffusionDrivenFlowSchematic}
\end{figure}

Figure \ref{fig:DiffusionDrivenFlowSchematic} sketches two coordinate systems for a tilted parallel-plate channel domain with an inclination angle $\theta$  relative to the horizontal plane, which  satisfies $0\leq \theta \leq \frac{\pi}{2}$. In this setup, $x_3$-direction is parallel to the direction of gravity. $\Omega=\left\{ y_{3}| y_3 \in [0,H] \right\}$ is the cross-section of the channel, where $H$ is the distance between plates. The longitudinal direction of the channel is denoted as the $y_1$-direction, with $y_{1}\in [-L,L]$, where $L$ may be either infinite or a finite number that is significantly larger than $H$. The relation between  the lab frame coordinates $(x_{1},x_{2},x_{3})$ and the coordinates $(y_{1},y_{2},y_{3})$ is 
\begin{equation}
\begin{aligned}
  \begin{bmatrix}
    y_{1}\\
    y_{3}\\
  \end{bmatrix}
  &=
  \begin{bmatrix}
    \cos \theta &\sin \theta \\
    -\sin \theta &\cos \theta\\
  \end{bmatrix}
  \begin{bmatrix}
    x_{1}\\
    x_{3}\\
  \end{bmatrix}, \quad y_{2}=x_{2}.
\end{aligned}
\end{equation}
In the $(y_{1},y_{2},y_{3})$ coordinate system, gravity acts along the direction $(-\sin \theta, 0, -\cos \theta)$. Experimental methods described in \cite{allshouse2010novel,heitz2005optimizing} provide feasible setups for this study. Another promising experimental configuration involves using temperature-stratified liquid gallium (\cite{braunsfurth1997free}).

  We consider a scalar $c$  responsible for fluid density stratification, referred to as the stratified  scalar in the subsequent context, such as temperature or solute concentration. The density is assumed to be a specified function of the stratified scalar.  For example, \cite{abaid2004internal} obtained the formula for the density of a sodium chloride solution as $\rho = 0.9971 + 0.00065S + 0.000322(25-T)$ g/cm$^{3}$, where salinity $S$ is measured in parts per thousand (ppt) and temperature $T$ is in degrees Celsius. In cases with a broader range of salinity, the relationship may exhibit nonlinearity.  \cite{hall1924densities}, through empirical data fitting, established  $\rho =0.997071+0.00070109 S+ 1.3268\times 10^{-7}S^{2}+3.535\times 10^{-10} S^3 $ g/cm$^{3}$  for  sodium chloride solutions at 25 degrees Celsius. The theoretical method presented in this work can be easily extended to cases with variable fluid viscosity. However, for simplicity, we assume the fluid dynamic viscosity is constant throughout the paper.

 The stratified  scalar  $c$ and the fluid flow $\mathbf{u}=(u_{1},u_{2},u_{3})$ in the $\mathbf{x}= (x_{1},x_{2},x_{3})$ direction satisfy the incompressible Navier-Stokes equation,
\begin{equation}\label{eq:NS dimensional 3D diffusion driven flow}
\begin{aligned}
  &\rho (c) \left(  \partial_{t}  u_{i}+  \mathbf{u}\cdot \nabla  u_{i} \right) = \mu \Delta u_{i} - \partial_{x_{i}} p- \rho (c) g \delta_{i3},\quad  \left. u_{i} \right|_{\mathbf{x}\in \partial \Omega }=0,\quad i=1,2,3,\\
&\nabla \cdot \mathbf{u}=0,\quad \partial_{t}c+ \mathbf{u}\cdot \nabla c= \kappa \Delta c,\quad \left. \partial_{\mathbf{n}}c \right|_{\mathbf{x}\in\partial \Omega }=0, \quad \left. c \right|_{t=0 }=c_{I} (\mathbf{x}). \\
\end{aligned}
\end{equation}
where $\mathbf{n}$ is the outward normal vector of the boundary, $\delta_{ij}$ is the Kronecker delta, $g$ (cm/s$^{2}$)  is the acceleration of gravity, $\rho$ $(\text{gram} \cdot \text{cm}^{-3})$ is the density, $\mu$ (gram$\cdot \text{cm}^{-1} \cdot \text{s}^{-1})$ is the dynamic viscosity,  $p$ (gram$\cdot \text{cm}^{-1}\cdot \text{s}^{-2})$ is the pressure and $\kappa$ (cm$^{2}/$s) is the molecular diffusivity of the stratified scalar.

It is convenient to consider the problem in $(y_{1},y_{2},y_{3})$ coordinate system. We denote $v_i$ as the velocity component along the $y_{i}$-direction. Since the initial condition and the boundary condition are independent of $y_2$, equation\eqref{eq:NS dimensional 3D diffusion driven flow} reduces to a two-dimensional problem
\begin{equation}\label{eq:NS dimensional 3D diffusion driven flow rotated coordinate}
  \begin{aligned}
&\rho \left(   \partial_{t}  v_{1}+ v_{1}\partial_{y_{1}} v_{1} + v_{3}\partial_{y_{3}} v_{1}\right) =\mu \Delta v_{1} - \partial_{y_{1}} p-   \sin \theta g \rho ,  \\
&\rho \left(   \partial_{t}  v_{3}+ v_{1}\partial_{y_{1}} v_{3} + v_{3}\partial_{y_{3}} v_{3}\right) =\mu \Delta v_{3} - \partial_{y_{3}} p-  \cos \theta g\rho ,  \\
&\partial_{t}c+  \mathbf{v} \cdot \nabla c= \kappa \Delta c, \quad \nabla \cdot \mathbf{v}=0,\quad \left. \mathbf{v} \right|_{\mathbf{y}\in \partial \Omega }=\mathbf{0}, \;\left. \partial_{\mathbf{n}}c \right|_{\mathbf{y}\in\partial \Omega }=0.\\
\end{aligned}
\end{equation}
In this study, we assume the initial density profile is a stable density stratification, where the density  decreases monotonically as the height increases, namely, $\partial_{y_{1}}\rho(\bar{c}_{I}) \leq 0$.  For convenience,  we use the overline to represent the cross-sectional average of a quantity and use the tilde to represent the fluctuation, for example, $\bar{c_{I}}(y_{1})= \int\limits_0^1 c_{I}(y_{1},y_{3}) \mathrm{d} y_{3}$.

\subsection{Nondimensionalization }
As the flow is driven by molecular diffusion, we select the diffusion time scale as the characteristic time, $\frac{H^{2}}{\kappa}$. Specifically, it refers to the time it takes for solute molecules to diffuse across the cross-sectional area of the channel domain. Utilizing the following change of variables for nondimensionalization,
\begin{equation}
\begin{aligned}
  &\rho_{0}\rho'=\rho , \quad c_{0}c'=c, \quad\frac{H^2}{\kappa} t' =t, \quad H\mathbf{y}'=\mathbf{y}, \quad U\mathbf{v}'=\mathbf{v}, \quad \frac{\mu U}{H} p' =p,
\end{aligned}
\end{equation}
Equation \eqref{eq:NS dimensional 3D diffusion driven flow rotated coordinate} becomes
\begin{equation}
  \begin{aligned}
 &\rho' \left(   \frac{U \kappa}{H^2}  \partial_{t'}  v_{1}'+ \frac{U^{2}}{H} v_{1}'\partial_{y_{1}'} v_{1}' + \frac{U^{2}}{H} v_{3}'\partial_{y_{3}'} v_{1}'\right) =\frac{ \mu U}{H^2\rho_{0}}  \Delta_{\mathbf{y}'} v_{1}' - \frac{\mu U}{H^2\rho_{0}}\partial_{y_{1}'} p'-   \sin \theta g \rho' ,  \\
    &\rho' \left(  \frac{U \kappa}{H^2}  \partial_{t'}  v_{3}'+ \frac{U^{2}}{H}  v_{1}'\partial_{y_{1}'} v_{3}' +\frac{U^{2}}{H} v_{3}'\partial_{y_{3}'} v_{3}'\right) =\frac{ \mu U}{H^2\rho_{0}}  \Delta_{\mathbf{y}'} v_{3}' - \frac{\mu U}{H^2\rho_{0}}\partial_{y_{3}'} p'-  \cos \theta g\rho' ,  \\
    &\frac{c_0\kappa}{H^2} \partial_{t'}c'+ \frac{U c_{0}}{H} \mathbf{v}' \cdot \nabla _{\mathbf{y}'}c'= \frac{\kappa c_{0}}{H^2} \Delta_{\mathbf{y}'}c',\quad \frac{c_{0}}{H} \left. \partial_{\mathbf{n}'}c' \right|_{\mathbf{y}'\in \partial \Omega'}=0.\\
\end{aligned}
\end{equation}
We can drop the primes without confusion and obtain the nondimensionalized version
\begin{equation}\label{eq:NS nondimensional 3D diffusion driven flow rotated coordinate}
  \begin{aligned}
&\rho \left( \frac{1}{ \mathrm{Sc}}   \partial_{t}  v_{1}+ \mathrm{Re} v_{1}\partial_{y_{1}} v_{1} + \mathrm{Re} v_{3}\partial_{y_{3}} v_{1}\right) =\Delta v_{1} - \partial_{y_{1}} p- \mathrm{Re} \mathrm{Ri}\rho  \sin \theta,  \\
&\rho \left( \frac{1}{ \mathrm{Sc}}  \partial_{t}  v_{3}+ \mathrm{Re} v_{1}\partial_{y_{1}} v_{3} +\mathrm{Re}  v_{3}\partial_{y_{3}} v_{3}\right) = \Delta v_{3} - \partial_{y_{3}} p- \mathrm{Re} \mathrm{Ri}\rho  \cos \theta,  \\
&\partial_{t}c+ \mathrm{Pe}  \mathbf{v} \cdot \nabla c= \Delta c,\\
\end{aligned}
\end{equation}
where the non-dimensional parameters are  P\'eclet number $\mathrm{Pe}= \frac{UH}{\kappa}$, Reynolds number $\mathrm{Re}= \frac{\rho_{0} H U}{\mu}$,  Richardson number $\mathrm{Ri}= \frac{g H}{U^{2}}$, and Schmidt number $\mathrm{Sc}= \frac{\mu}{\rho_{0} \kappa}=\frac{\mathrm{Pe}}{\mathrm{Re}}$. If the scalar field is the temperature field, then $\kappa$ is the thermal diffusivity and $\mathrm{Pr}=\frac{\mu}{\rho_{0} \kappa}=\frac{\mathrm{Pe}}{\mathrm{Re}}$ is the Prandtl number.

We proceed by examining a combination of experimental physical parameters, which can provide us with the order of magnitude of the non-dimensional parameters and assist in our perturbation analysis. The previous experiments primarily focused on diffusion in linearly stratified fluids. It's feasible to adapt these setups for experiments involving nonlinearly stratified fluids, making the parameters therein useful as reference points. For instance, in \cite{camassa2019first}, a linear density stratification of the fluid was achieved using sodium chloride.
The relevant physical parameters in this context are as follows:  $g=980$  cm/s$^{2}$, $\mu_{0}=0.008903$ gram/(cm.s), $\kappa= 1.5\times10^{-5}$ cm$^{2}$/s (\cite{vitagliano1956diffusion}), $\rho_{0}= 0.9971$ gram/cm$^{3}$, the vertical density gradient $\partial_{x_{3}}\rho$ is a constant $0.007$ gram/$cm^{4}$. The scaling relation for the characteristic velocity and the physical parameters varies depending on the boundary geometries. For a linear stratified fluid, the characteristic velocity and characteristic boundary layer thickness of steady diffusion-driven flow in a parallel-plate channel can be calculated using the formula from (\cite{phillips1970flows,heitz2005optimizing}) as follows
\begin{equation}
\begin{aligned}
&U =2 \kappa  \cot (\theta ) \left( \frac{   g \sin (\theta) \partial_{x_{3}}\rho}{4 \kappa  \mu } \right)^{\frac{1}{4}} ,\quad H_{b}=\left( \frac{  g \sin (\theta )\partial_{x_{3}}\rho }{4 \kappa  \mu } \right)^{-\frac{1}{4}}.
\end{aligned}
\end{equation}
Using these formulas, we obtain the values $U =0.00164684$ cm/s, $H_{b}= 0.0182167$ cm for $\theta=\frac{\pi}{4}$. When the gap thickness between two walls is $H=0.1$ cm, this leads to the following non-dimensional parameters:
\begin{equation}\label{eq:non dimensional parameter1}
  \mathrm{Re}=0.0184512,\quad  \mathrm{Pe}=10.9799, \quad \mathrm{Sc}=595.076, \quad \mathrm{Re} \mathrm{Ri}=666613.
\end{equation}
It's evident that the Reynolds number is small, indicating that viscous effects dominate the flow, and the gravity term is significant in the governing equation.

Last, with the above parameters and  $c_{0}=1$ mol/kg, the relation between $\rho$ and $c$ can be expressed as
\begin{equation}\label{eq:density concentration}
\begin{aligned}
&\rho (c)=1+0.0410921 c+ 0.000454464 c^2+ 0.000070761 c^3.
\end{aligned}
\end{equation}

\section{Asymptotic analysis and numerical simulation}
\label{sec:Flow for slowly varying density profile}
In this section, our goal is to derive approximations for the velocity and density fields and validate them through numerical simulations.

The hydrostatic body force terms in the governing equation contribute only to the hydrostatic pressure field. Since they do not influence the velocity and stratified scalar field, we can absorb them into a modified pressure. For convenience,  we define $p=p_{0}+\tilde{p}$, where
\begin{equation}
\begin{aligned}
&p_{0}=- \mathrm{Re} \mathrm{Ri}   \left(\sin \theta  \int\limits_{0}^{y_{1}}\rho(\bar{c} (y_{1})) \mathrm{d}y_{1}+ \cos \theta(y_{3}-\frac{1}{2})\rho(\bar{c}) \right), \\
&\partial_{y_{1}}p_{0}=- \mathrm{Re} \mathrm{Ri}    \left(\sin \theta  \rho( \bar{c}) + \cos \theta(y_{3}-\frac{1}{2})\partial_{y_{1}}\rho (\bar{c}) \right) ,\; \partial_{y_{3}}p_{0}=- \mathrm{Re} \mathrm{Ri}   \cos \theta \rho(\bar{c}). \\
\end{aligned}
\end{equation}

Then the governing equation \eqref{eq:NS nondimensional 3D diffusion driven flow rotated coordinate}   becomes 
\begin{equation}\label{eq:NS nondimensional 2D diffusion driven flow Boussinesq}
  \begin{aligned}
    &\rho  \left( \frac{1}{ \mathrm{Sc}}   \partial_{t}  v_{1}+ \mathrm{Re} v_{1}\partial_{y_{1}} v_{1} + \mathrm{Re} v_{3}\partial_{y_{3}} v_{1}\right) \\
    &\hspace{2cm}=\Delta v_{1} - \partial_{y_{1}} \tilde{p}- \mathrm{Re} \mathrm{Ri} \left( \sin \theta  (\rho - \rho(\bar{c})) -\cos \theta(y_{3}-\frac{1}{2})\partial_{y_{1}}\rho (\bar{c})\right),  \\
&\rho \left( \frac{1}{ \mathrm{Sc}}  \partial_{t}  v_{3}+ \mathrm{Re} v_{1}\partial_{y_{1}} v_{3} +\mathrm{Re}  v_{3}\partial_{y_{3}} v_{3}\right) = \Delta v_{3} - \partial_{y_{3}} \tilde{p}- \mathrm{Re} \mathrm{Ri}   \cos \theta (\rho - \rho(\bar{c})),  \\
&\partial_{t}c+ \mathrm{Pe} \left( v_{1}\partial_{y_{1}}c+v_{3}\partial_{y_{3}}c \right) = \Delta c,\quad \partial_{y_{1}}v_{1}+\partial_{y_{3}}v_{3}=0.
\end{aligned}
\end{equation}

\subsection{Expansions for slow varying density profile}

To address the scenario where the longitudinal length scale of the channel domain significantly exceeds the transverse length scale, our objective is to develop a simplified model that relies solely on the longitudinal variable of the channel domain. To achieve this, we adopt the following ansatz for the velocity and density, where  
\begin{equation}\label{eq:asymptotic expansion ansatz}
\begin{aligned}
&v_{1} (y_{1},y_{3},t) =v_{1,0}(y_{3},t, \bar{c})+v_{1,1} \left(y_{3},t, \partial_{y_{1}}\bar{c} \right)\partial_{y_{1}}\bar{c}+v_{1,2} \left(y_{3},t, \partial_{y_{1}}\bar{c},\partial_{y_{1}}^{2}\bar{c} \right) \partial_{y_{1}}^{2}\bar{c}+\hdots, \\
&v_{3}(y_{1},y_{3},t)=v_{3,0}(y_{3},t,\bar{c})+v_{3,1} \left(y_{3},t,  \partial_{y_{1}}\bar{c} \right)\partial_{y_{1}}\bar{c}+v_{3,2} \left( y_{3},t, \partial_{y_{1}}\bar{c},\partial_{y_{1}}^{2}\bar{c} \right) \partial_{y_{1}}^{2}\bar{c}+\hdots, \\
&c(y_{1},y_{3},t)=\bar{c}(y_{1},t)+ c_{1} \left(y_{3},t,  \partial_{y_{1}}\bar{c} \right)\partial_{y_{1}}\bar{c}+c_{2} \left( y_{3},t, \partial_{y_{1}}\bar{c},\partial_{y_{1}}^{2}\bar{c} \right) \partial_{y_{1}}^{2}\bar{c}+\hdots,\\
&p(y_{1},y_{3},t)=p_{0}(y_{1},y_{3},t)+ p_{1} \left(y_{3},t,  \partial_{y_{1}}\bar{c} \right)\partial_{y_{1}}\bar{c}+p_{2} \left( y_{3},t, \partial_{y_{1}}\bar{c},\partial_{y_{1}}^{2}\bar{c} \right) \partial_{y_{1}}^{2}\bar{c}+\hdots.\\
\end{aligned}
\end{equation}
In this expansion, $v_{i,1},c_{1},p_{1}$ are functions of $\partial_{y_{1}}\bar{c}$. $v_{j,2},c_{2},p_{2}$ are functions of $\partial_{y_{1}}\bar{c}$ and $\partial_{y_{1}}^{2}\bar{c}$. The $j$-th term in the expansion is contingent on the derivative of $\bar{c}$ up to the $j$th order. By definition, $\partial_{y_{1}}\bar{c}$ is independent of $y_{3}$ and is solely a function of $y_{1}$ and $t$. For the specific problem we investigate here,  both $v_{i,j}$ and $c_{i}$ are functions of $y_{3}$ and $t$, while their dependence on $y_{1}$ is captured through $\partial_{y_{1}}\bar{c}$.   In more general cases, such as the domain with non-flat boundary, $v_{i,j}$ and $c_{i}$ may depend on $y_{1}$.

In our analysis, we consider the derivative of the averaged stratified scalar,  $\partial_{y_{1}}^{n}\bar{c}$, as a small parameter within the framework of standard asymptotic calculations. Additionally, we assume that higher-order derivatives exhibit smaller magnitudes compared to lower-order derivatives. This assumption is reasonable for systems where diffusion dominates.

To provide an intuitive justification, let's assume that flow effects are negligible, and diffusion is the dominant process in the system. Under these conditions, the scalar field exhibits a self-similarity solution represented as $\bar{c} = \frac{1}{2} \text{erfc}\left(\frac{y_{1}}{2 \sqrt{t}}\right)$, where $\mathrm{erfc}(z) = \frac{2}{\sqrt{\pi}}\int_z^\infty e^{-t^2} dt$. The derivatives of this solution are as follows:
\begin{equation}
\begin{aligned}
  &\partial_{y_{1}}\bar{c} = -\frac{e^{-\frac{y_{1}^2}{4 t}}}{2 \sqrt{\pi} \sqrt{t}},\quad \partial_{y_{1}}^{2}\bar{c} =- \frac{y_{1} e^{-\frac{y_{1}^2}{4 t}}}{4 \sqrt{\pi} t^{3/2}},\\
  &\partial_{y_{1}}^{n}\bar{c}= -\frac{1}{\sqrt{\pi }}\left(-\frac{1}{2 \sqrt{t}}\right)^n  H_{n-1}\left(\frac{y_{1}}{2 \sqrt{t}}\right) e^{-\left(\frac{y_{1}}{2 \sqrt{t}}\right)^2},
\end{aligned}
\end{equation}
where $H_{n}$ is the Hermite polynomial of degree $n$. In general, as $t$ approaches infinity, we have $\partial_{y_{1}}^{n}\bar{c} = \mathcal{O}(t^{-n+\frac{1}{2}})$. Consequently, as $t\rightarrow \infty$, we observe the hierarchy $\bar{c} \gg \partial_{y_{1}}\bar{c} \gg \partial_{y_{1}}^{2}\bar{c} \gg \hdots$. Due to the orthogonality of the Hermite polynomial, $\partial_{y_{1}}^{n}\bar{c}$ forms a good basis for approximating the function.  In addition, it's important to note that in the presence of a diffusion-driven flow, as we will demonstrate in the following sections, the scaling relationship can differ. For instance, in some parameter limit, the self-similarity variable in the solution can be   $y_{1}t^{-\frac{1}{4}}$ for finite $t$. However, in this case, the derivatives of the averaged density still tend to be zero, and the higher-order derivatives become much smaller than the lower-order derivatives as $t\rightarrow \infty$.

The similar form of expansion \eqref{eq:asymptotic expansion ansatz}  has found applications in various fields, including the modeling of chromatograph and reactors (\cite{balakotaiah1995dispersion}), thin film fluid flows (\cite{van1987slow,roberts1996low,roberts2006accurate}), and shear dispersion of passive scalars (\cite{gill1967note,young1991shear,mercer1990centre,ding2022determinism}). A more rigorous foundation for this expansion can be established through center manifold theory (\cite{carr1979applications,aulbach1996integral, aulbach1999invariant,roberts1988application,roberts2014model,roberts2015macroscale}). For in-depth discussions on constructing center manifolds, we refer interested readers to the cited literature, and we will not delve into the details here.  

Another possible asymptotic expansion approach is a power series expansion involving a small parameter, denoted as $\epsilon$ , such as $c = \sum\limits_{i=0}^{\infty}a_{i}\epsilon^{i}$. This approach is commonly used in multiscale analysis (\cite{kondic2003instabilities,pavliotis2008multiscale,wu2014approach,ding2023shear}).    In this context, we highlight two advantages of the ansatz provided by Equation \eqref{eq:asymptotic expansion ansatz} over the classical power series expansion for the specific problem addressed in this study.

First, in the classical ansatz, the coefficients remain independent of the small parameter $\epsilon$. The quantity is expanded as a linear combination of $\epsilon$ powers. In contrast, in Equation \eqref{eq:asymptotic expansion ansatz}, the coefficients depend on small quantities, namely,  the derivatives of the averaged scalar field. This nonlinear dependency enables us to achieve a more accurate approximation using fewer terms.

Second, when applying the standard multiscale analysis method to this problem, the results depend on the scaling relation between parameters. In the thin film limit, as we will discuss in Section \ref{sec:power series expansion}, the scalar field can be approximated as $c=c_{0} (y_{1},t) +\epsilon c_{2}(y_{1},y_{3},t)$, where $c_{0}$ is the solution to a diffusion equation with a diffusivity of 1.  In this scaling relation, the diffusion process is dominant, and the diffusion-driven flow does not have a first-order contribution. For cases where the diffusion-driven flow significantly enhances the dispersion of the stratified agent, an asymptotic analysis using a different scaling relation for physical parameters become necessary.  In some cases, selecting the most appropriate characteristic parameter can be challenging. Additionally, since certain scales in the problem are time-dependent, the choice of scale may lead to time-dependent small parameters, which can complicate the development of a model that is uniformly valid across a wide range of parameter regimes.

In Section \ref{sec:power series expansion}, we will provide detailed explanations of the thin film approximation and discuss the differences between these two approaches.

\subsection{Leading order term in Asymptotic expansion }
The expansion \eqref{eq:asymptotic expansion ansatz} suggests that the deviation of the stratified  scalar from its cross-sectional average is small. Consequently, this motivates us to examine the evolution equation for the cross-sectional averaged stratified  scalar $\bar{c}$.  Taking the cross sectional average on both side of the advection-diffusion equation \eqref{eq:NS nondimensional 2D diffusion driven flow Boussinesq} and utilizing the incompressibility condition and non-flux boundary condition yields
\begin{equation}\label{eq:averaged equaiton for density}
\begin{aligned}
&\partial_{t} \bar{c}+ \mathrm{Pe}\partial_{y_{1}} \overline{v_{1}c}= \partial_{y_{1}}^{2}\bar{c}.
\end{aligned}
\end{equation}
Therefore, to obtain the leading order approximation for the cross-sectional averaged stratified  scalar, we need to compute $v_{1,0}$, $v_{1,1}$ and $c_{1}$ in the expansion \eqref{eq:asymptotic expansion ansatz}.

There are some observations that can  simplify the calculation of asymptotic expansion. Since the flow is induced via the density  gradient, the fluid flow vanishes as  $\partial_{y_{1}}\bar{c}$ vanishes. Therefore, $v_{1,0}=v_{3,0}=0$.  Substituting the expansion \eqref{eq:asymptotic expansion ansatz} into the continuity equation and noticing that $\partial_{y_{1}} \left( v_{1,1}\partial_{y_{1}}\bar{c} \right)= \mathcal{O}\left( \partial_{y_{1}}^{2} \bar{c}\right)$, we obtain
\begin{equation}
\begin{aligned}
&\partial_{y_{1}}v_{1,0}+ \partial_{y_{3}}v_{3,1}=0.
\end{aligned}
\end{equation}
Then to satisfy the no-slip boundary condition,  $v_{3,1}$ must be $0$.

 Collecting the terms that are comparable to $\partial_{y_{1}}\bar{c}$ yields the following equation 
\begin{equation}\label{eq:diffusion-driven flow parallel-plate channel domain axillary problem1}
\begin{aligned}
&\frac{1}{\mathrm{Sc}}\rho(\bar{c})\partial_{t}v_{1,1}= \partial_{y_{3}}^2v_{1,1}- \mathrm{Re} \mathrm{Ri} \partial_{c}\rho (\bar{c})  \left( c_{1} \sin \theta     -\cos \theta(y_{3}-\frac{1}{2})\right),  \\
&\partial_{t}c_{1}+ \mathrm{Pe} v_{1,1} \partial_{y_{1}}\bar{c}=\partial_{y_{3}}^{2} c_{1}, \quad \left. \partial_{y_3}c_{1} \right|_{y_{3}=0,1 }=0,\quad \left. v_{1,1} \right|_{y_{3}=0,1 }=0. \\
\end{aligned}
\end{equation}
Previous studies (\cite{kistovich1993structure, ding2023dispersion}) have uncovered fascinating dynamics in the time-dependent solution, particularly during short time scales when the initial velocity field is zero. However, these transient dynamics decay exponentially, and the solution converges to a steady-state either at the diffusion time scale or the viscosity time scale, depending on which of the two is larger. In many cases, the diffusion time scale significantly exceeds or is comparable to the viscosity time scale. For instance, in solute-liquid systems, the diffusivity of the solute molecule typically falls within the range of $10^{-5}$ cm$^{2}$/s, while the kinematic viscosity of the liquid is around $10^{-2}$ cm$^{2}$/s, resulting in a Schmidt number ($\mathrm{Sc}$) of $10^{3}$.  In the case of temperature-stratified systems, the thermal diffusivity typically registers at approximately $10^{-3}$ cm$^{2}$/s, with $\mathrm{Pr}=10$. In both cases,  the diffusion time scale is much larger than the viscosity time scale. Therefore, given our specific interest in approximations at the diffusion time scale and to streamline our analysis without compromising accuracy, our primary focus will be on the steady-state solution of the aforementioned equation. By neglecting the time derivatives in equation \eqref{eq:diffusion-driven flow parallel-plate channel domain axillary problem1}, we arrive at the following non-homogeneous linear system for analysis:
\begin{equation}\label{eq:diffusion-driven flow parallel-plate channel domain simplified}
\begin{aligned}
&0= \partial_{y_{3}}^2v_{1,1}- \mathrm{Re}\mathrm{Ri}\partial_{c}\rho (\bar{c})  \left(\sin \theta  c_{1}     -\cos \theta(y_{3}-\frac{1}{2})\right), \quad \left. v_{1,1} \right|_{y_{3}=0,1 }=0, \\
&\mathrm{Pe} v_{1,1} \partial_{y_{1}}\bar{c}=\partial_{y_{3}}^{2} c_{1}, \quad \left. \partial_{y_3}c_{1} \right|_{y_{3}=0,1 }=0. \\
\end{aligned}
\end{equation}
Differencing the above equation twice with respect to $y_{3}$ yields 
\begin{equation}
\begin{aligned}
&0= \partial_{y_{3}}^4v_{1,1}- \mathrm{Re} \mathrm{Ri}  \sin \theta \partial_{c}\rho  \partial_{y_{3}}^{2} c_{1}, \; \mathrm{Pe}  \partial_{y_{1}}\bar{c} \partial_{y_{3}}^{2}v_{1,1}=\partial_{y_{3}}^{4} c_{1}, \; \left. \partial_{y_{3}}^{3}v_{1,1} \right|_{y_{3}=0,1 }= -\mathrm{Re} \mathrm{Ri}  \partial_{c}\rho (\bar{c})  \cos \theta.  \\
\end{aligned}
\end{equation}
Then we can decouple the scalar field and velocity as 
\begin{equation}
\begin{aligned}
&0=\partial_{y_{3}}^4v_{1,1}  -\mathrm{Re} \mathrm{Pe} \mathrm{Ri}\sin \theta  \partial_{y_{1}}\rho(\bar{c}) v_{1,1}, \quad \left. v_{1,1} \right|_{y_{3}=0,1 }=0, \quad \left. \partial_{y_{3}}^{3}v_{1,1} \right|_{y_{3}=0,1 }= -\mathrm{Re} \mathrm{Ri}  \partial_{c}\rho (\bar{c})  \cos \theta,   \\
&\partial_{y_{3}}^{4} c_{1}=  \mathrm{Re} \mathrm{Pe}  \mathrm{Ri}  \partial_{y_{1}}\rho(\bar{c}) \left( \sin \theta c_{1}  -\cos \theta(y_{3}-\frac{1}{2}) \right) , \quad \left. \partial_{y_3}c_{1} \right|_{y_{3}=0,1 }=0, \quad \left. \partial_{y_2}^{2}c_{1} \right|_{y_{3}=0,1 }=0,
\end{aligned}
\end{equation}
where we have used $\partial_{y_{1}}\rho (\bar{c})=\partial_{c}\rho (\bar{c}) \partial_{y_{1}} \bar{c}  $ to simplify the equation.  We can solve the equation and express $v_{1,1}$, $c_{1}$ in terms of $\partial_{y_{1}}\bar{c}$
\begin{equation}\label{eq:first order approximation density}
\begin{aligned}
  &v_{1,1}= \frac{2 \gamma  \cot (\theta )}{\mathrm{Pe}\partial_{y_{1}} \bar{c}} \frac{ \sin (\gamma  y_{3}) \sinh (\gamma (1-  y_{3}))-\sin (\gamma (1- y_{3})) \sinh (\gamma  y_{3}) }{ \sin (\gamma )+\sinh (\gamma )},\\
 &c_{1}=\cot \theta  \left( y_{3}- \frac{1}{2}-\frac{  \cos (\gamma(1- y_{3})) \cosh (\gamma y_{3})-\cos (\gamma y_{3}) \cosh (\gamma(1- y_{3}))}{\gamma (\sin (\gamma)+\sinh (\gamma))} \right),\\
\end{aligned}
\end{equation}
where $\gamma= \frac{ 1}{\sqrt{2}}\left( -\mathrm{Re} \mathrm{Pe}  \mathrm{Ri} \sin \theta \partial_{y_{1}}\rho  (\bar{c})\right)^{\frac{1}{4}}$. Recall our assumption of the stable density stratification, $ \partial_{y_{1}}\rho  (\bar{c})\leq 0$, which implies that $\gamma$ is a real nonnegative number. In the case of linear density stratification, 
the equation above aligns with the steady solution previously presented in (\cite{phillips1970flows,heitz2005optimizing}).

\begin{figure}
  \centering
    \subfigure[]{
    \includegraphics[width=0.46\linewidth]{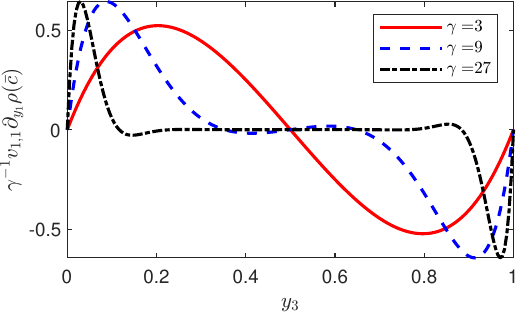}
  }
    \subfigure[]{
    \includegraphics[width=0.46\linewidth]{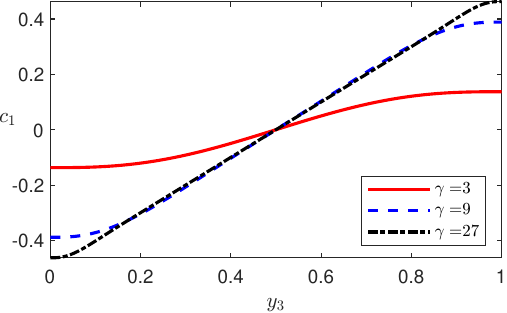}
  }

  \hfill
  \caption[]
  {(a) Normalized leading order approximation of the longitudinal velocity $\frac{v_{1,1}\partial_{y_{1}}\rho (\bar{c})}{\gamma}$ for various parameter $\gamma$. (b) $c_{1}$ for different $\gamma$. }
  \label{fig:DiffusionDrivenFlowSteadySolution}
\end{figure}

As depicted in Figure \ref{fig:DiffusionDrivenFlowSteadySolution}, both the normalized velocity $v_{1} \gamma^{-1}\approx v_{1,1}\partial_{y_{1}}\bar{c}\gamma^{-1}$ and $c_{1}$ are tightly confined within a narrow region near the boundary, especially when $\gamma$ is large. The velocity is positive near $y_{3}=0$ indicating a upwelling flow near the upward facing boundary. Notably, the normalized velocity exhibits nearly uniform magnitude across different values of $\gamma$. Thus, $\gamma^{-1}$ serves as an effective indicator of the boundary layer's thickness, and $\gamma$ can be considered the characteristic velocity of the system. Moreover, the functions illustrated in Figure \ref{fig:DiffusionDrivenFlowSteadySolution} display an inherent odd symmetry about $y_{3}=\frac{1}{2}$. This symmetry suggests that the leading-order approximations for both the velocity and density fields also possess odd symmetry with respect to $y_{3}=\frac{1}{2}$. As a result, their cross-sectional averages inherently amount to zero.

As $\partial_{y_{1}}\bar{c}\rightarrow 0$, we have
\begin{equation}\label{eq:velocity low density gradient}
\begin{aligned}
&  v_{1,1}=-\frac{\text{Re} \mathrm{Ri} y_{3}(y_{3}-1) (2 y_{3}-1) \cos (\theta )\partial_{c}\rho}{12 }\\
  &-\frac{  \text{Pe} \text{Re}^2  \mathrm{Ri}^{2}y_{3} \left(2 (y_{3} (2 y_{3}-7)+7) y_{3}^4-7 y_{3}+3\right) \sin (2 \theta ) \partial_{y_{1}}\rho\partial_{c}\rho}{40320}+\mathcal{O}\left( \left( \partial_{y_{1}}\bar{c} \right)^{2}\right), \\
\end{aligned}
\end{equation}
and
\begin{equation}\label{eq:rho low density gradient}
\begin{aligned}
c_{1}=&\frac{\text{Pe} \text{Re} \mathrm{Ri} \left(1-2 y_{3}^3 (3 y_{3} (2 y_{3}-5)+10)\right) \cos (\theta ) \partial_{y_{1}}\rho}{1440}+\mathcal{O}\left( \left( \partial_{y_{1}}\bar{c} \right)^{2}\right). \\
\end{aligned}
\end{equation}
Therefore $v_{1,1}\sim \mathcal{O} (1)$, but $c_{1}\sim \mathcal{O} (\partial_{y_{1}}\bar{c})$. The leading order approximation of the longitudinal velocity and the density field are given by 
$v_{1}\sim v_{1,1}\partial_{y_{1}}\bar{c} = \mathcal{O} (\partial_{y_{1}}\bar{c})$ and $c \sim c_{1}  \partial_{y_{1}}\bar{c}= \mathcal{O} \left(  \left( \partial_{y_{1}}\bar{c} \right)^{2} \right)$, which confirms the ansatz of the asymptotic expansion \eqref{eq:asymptotic expansion ansatz}.

Substituting the expansion of the velocity field and scalar field to equation \eqref{eq:averaged equaiton for density} and noticing that $\bar{v}_{1,1}=0$, the leading order approximation reads
\begin{equation}\label{eq:effective equation 0}
\begin{aligned}
  &\partial_{t}\bar{c}+ \mathrm{Pe}  \partial_{y_{1}}  \left(   \overline{v_{1,1}c_{1}}  \left( \partial_{y_{1}}\bar{c} \right)^{2}  \right) = \partial_{y_{1}}^{2}\bar{c}.\\
\end{aligned}
\end{equation}
Computing the average yields
\begin{equation}
\begin{aligned}
 &  \overline{v_{1,1}c_{1}}=\frac{-\cot ^2(\theta ) }{  \text{Pe} \partial_{y_{1}}\bar{c}} \left( \frac{\sin (\gamma ) \sinh (\gamma )}{(\sin (\gamma )+\sinh (\gamma ))^2}+\frac{5 (\cos (\gamma )-\cosh (\gamma ))}{2 \gamma  (\sin (\gamma )+\sinh (\gamma ))}+1  \right).\\
\end{aligned}
\end{equation}
The effective equation \ref{eq:effective equation 0} for the averaged stratified  scalar becomes
\begin{equation}\label{eq:effective equation}
\begin{aligned}
  &\partial_{t}\bar{c}= \partial_{y_{1}}  \left( \kappa_{\mathrm{eff}} \partial_{y_{1}}\bar{c} \right),\\
  &\kappa_{\mathrm{eff}}=1 +\cot ^2(\theta ) \left( \frac{\sin (\gamma ) \sinh (\gamma )}{(\sin (\gamma )+\sinh (\gamma ))^2}+\frac{5 (\cos (\gamma )-\cosh (\gamma ))}{2 \gamma  (\sin (\gamma )+\sinh (\gamma ))}+1 \right)  .\\
 \end{aligned}
\end{equation}
This equation can be considered as an generalization of the diffusion equation with replacing the constant diffusion coefficient by a positive definite function that depends on the derivative of the solution. This equation belongs to the family of equation that takes the form $\partial_{t}w=f (\partial_{x}w) \partial_{x}^{2}w$, which occurs in the nonlinear theory of flows in porous media and also governs the motion of a nonlinear viscoplastic medium (See page 342 in \cite{polyanin2012handbook}).

To complete the leading order approximation for the whole system, we next consider the approximation for the transverse velocity and pressure. Once the leading order approximation the longitudinal velocity is known, we can calculate the transverse velocity via the  continuity equation. Substituting the  expansion \eqref{eq:asymptotic expansion ansatz} into the continuity equation and collecting the terms that are comparable to $\partial_{y_{1}}^{2}\bar{c}$ yields
\begin{equation}
\begin{aligned}
&\partial_{y_{1}}\left( v_{1,1} \partial_{y_{1}}\bar{c} \right)+\partial_{y_{3}}v_{3,2}\partial_{y_{1}}^{2}\bar{c}=0. \\
\end{aligned}
\end{equation}
Using no-slip boundary condition and  the fact that $\bar{v}_{1,1}=0$ yields the expression of $v_{3,2}$:
\begin{equation}
\begin{aligned}
&v_{3,2}= -\frac{ \int\limits_0^{y_{3}} \partial_{y_{1}}\left( v_{1,1} \partial_{y_{1}}\bar{c} \right)\mathrm{d} y_{3} }{\partial_{y_{1}}^{2} \bar{c}}\\
&=\frac{\cot (\theta )\gamma }
{2\text{Pe} (\sin (\gamma )+\sinh (\gamma ))^2 \partial_{y_{1}}\rho } \left( \sin (\gamma(1 - y)) \sinh (\gamma  y) (y \sin (\gamma )+(y-1) \sinh (\gamma ))\right.\\
  &\hspace{4cm}\left. -\sin (\gamma  y) \sinh (\gamma (1- y)) \left( (y-1) \sin (\gamma )+y \sinh (\gamma ) \right) \right).
\end{aligned}
\end{equation}

In the limit $\partial_{y_{1}}\bar{c}\rightarrow 0$, we have
\begin{equation}\label{eq:first order approximation v3}
\begin{aligned}
  v_{3,2}=&\frac{1}{24} \text{Re} \text{Ri} (y_{3}-1)^2 y_{3}^2 \cos (\theta )\\
  &+\frac{\text{Pe} \text{Re}^2 \text{Ri}^2 y_{3}^2 \left( y_{3}-1 \right)^{2} \left(3 y_{3}^4-6 y_{3}^3-y_{3}^2+4 y_{3}+9\right) \sin (2 \theta ) \partial_{y_{1}}\bar{\rho}}{120960}+ \mathcal{O} \left( \left( \partial_{y_{1}}\bar{c} \right)^{2} \right).
\end{aligned}
\end{equation}
Therefore, $v_{3} \sim v_{3,2}\partial_{y_{1}}^{2}\bar{c} = \mathcal{O} (\partial_{y_{1}}^{2} \bar{c})$, consistent with the assumption regarding the order of magnitude in the expansion given by \eqref{eq:asymptotic expansion ansatz}.

$p_{1}$ in the expansion of pressure provided in equation \eqref{eq:asymptotic expansion ansatz} satisfies
\begin{equation}
\begin{aligned}
&-\partial_{y_{3}}p_{1}  - \mathrm{Re} \mathrm{Ri}   \cos \theta c_{1}\partial_{c}\rho(\bar{c}) =0.
\end{aligned}
\end{equation}
The solution is
\begin{equation}\label{eq:first order approximation pressure}
\begin{aligned}
  &p_{1}= -\cot (\theta ) \text{cos}(\theta )\mathrm{Ri}\mathrm{Re}\partial_{c}\rho \left(\frac{1}{12}+  \frac{1}{2} (y-1) y+  \right. \\
&\left. \frac{\sin (\gamma -\gamma  y) \cosh (\gamma  y)-\cos (\gamma -\gamma  y) \sinh (\gamma  y)-\cos (\gamma  y) \sinh (\gamma -\gamma  y)+\sin (\gamma  y) \cosh (\gamma -\gamma  y)}{2 \gamma ^2 (\sin (\gamma )+\sinh (\gamma ))}   \right).
\end{aligned}
\end{equation}

In the limit $\partial_{y_{1}}\bar{c}\rightarrow 0$, we have
\begin{equation}
\begin{aligned}
  p_{1}=& \frac{ \cos^{2} \theta \text{Pe} \text{Re}^{2} \text{Ri}^{2} \left(28 y_{3}^6-84 y_{3}^5+70 y_{3}^4-14 y_{3}+3\right) \partial_{y_{1}}\rho \partial_{c}\rho}{20160} +\mathcal{O} \left( \left( \partial_{y_{1}}\bar{c} \right)^{2} \right).
\end{aligned}
\end{equation}

So far, we have derived the leading-order approximation for the entire system. The numerical simulations in the next section demonstrate that the leading-order approximation has decent accuracy in many scenarios. However, in cases of highly nonlinear scalar distribution, or where the density depends highly nonlinearly on the stratified scalar, higher-order terms are required to achieve desirable accuracy. Similar cases can be found in previous literature. For example, \cite{grayer2021stably} studied a stably stratified square cavity subjected to horizontal oscillations, which despite not being purely diffusion-driven, showing that approximations become less regular as viscosity decreases and how higher-order terms are needed.

\subsection{Numerical simulation}
\label{sec:Numerical simulation}
In this subsection, we perform simulations of the full governing equation \eqref{eq:NS nondimensional 3D diffusion driven flow rotated coordinate} to demonstrate both the accuracy and validity of our asymptotic approximation and the effective equation \eqref{eq:effective equation}. The details of the numerical methods are documented in appendix \ref{sec:appendix numerical method}.
The computational domain is defined as  $\left\{ (y_{1},y_{3}) | y_{1} \in [-5,5], y_{3}\in [0,1]\right\}$. The no-slip boundary condition is imposed for the velocity field, the no-flux boundary condition is imposed for the stratified scalar.  The density and stratified scalar relation \eqref{eq:density concentration} is used for all test cases in this section.

In the first numerical simulation,  the initial condition is 
\begin{equation}\label{eq:initial condition 1}
\begin{aligned}
&c_{I} = \frac{1}{2}\mathrm{erfc}(y_{1}) -\frac{e^{-x^2}}{\sqrt{\pi }} \cos \theta \left( y_{3}-\frac{1}{2} \right) ,\quad v_{1}=v_{3}=0.
\end{aligned}
\end{equation}
and the dimensionless parameters are
\begin{equation}\label{eq:parameters 1}
\mathrm{Re}=1,\quad \mathrm{Ri}\mathrm{Re}=12500,\quad   \mathrm{Pe}=40, \quad \mathrm{Sc}=1,\quad  \theta=\frac{\pi}{4}. 
\end{equation}

\begin{figure}
  \centering
      \subfigure[]{
    \includegraphics[width=0.46\linewidth]{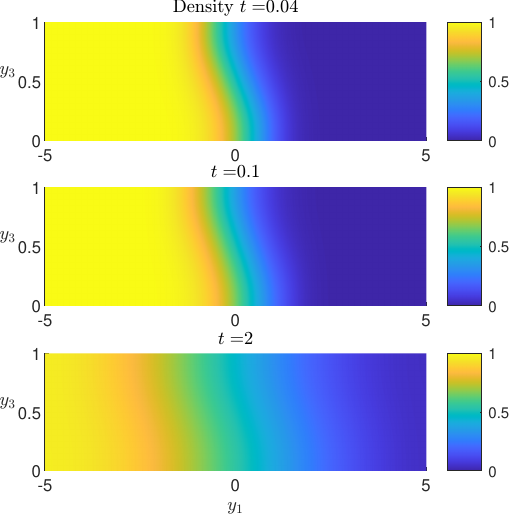}
  }
  \subfigure[]{
    \includegraphics[width=0.46\linewidth]{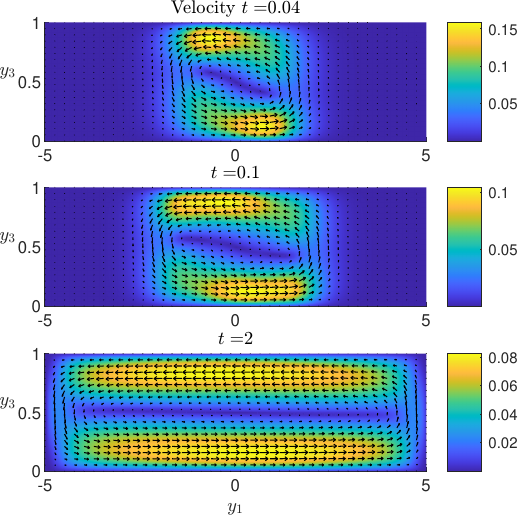}
  }

  \hfill
  \caption[]
  {    In the left panel, we display the density field at three time instances: $t=0.04$ (top), $t=0.1$ (middle), and $t=2$ (bottom). The right panel shows a pseudocolor plot illustrating the magnitude of the velocity field at the same time points.  The initial condition and dimensionless parameter are provided in equation \eqref{eq:initial condition 1} and \eqref{eq:parameters 1}, respectively.
  }
  \label{fig:FullSimulationtVelDensity}
\end{figure}
The stratified scalar and velocity at $t=0.04$ are depicted in the first row of Figure \ref{fig:FullSimulationtVelDensity} (a) and (b), respectively. Since density gradient is large around $y_{1}=0$, the flow is also localized in that regime. The third row of Figure \ref{fig:FullSimulationtVelDensity} displays the density field and velocity field obtained by the simulation at $t=2$. As time increases,  the density field becomes smoother and a large fluid circulation is formed in this whole domain.

\begin{figure}
  \centering
      \subfigure[]{
    \includegraphics[width=0.46\linewidth]{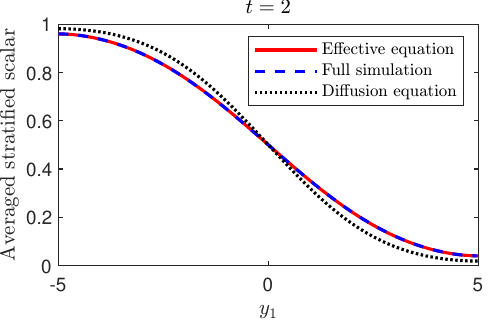}
  }
  \subfigure[]{
    \includegraphics[width=0.46\linewidth]{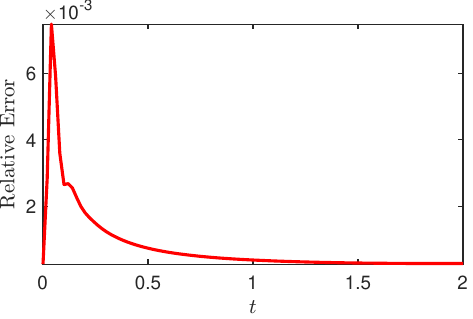}
  }
  \hfill
  \caption[]
  {In the left panel, we present the solution of the effective equation \eqref{eq:effective equation} (represented by the red solid curve), the cross-sectional averaged stratified  scalar derived from simulating the complete governing equation (illustrated by the blue dashed curve), and the solution to the diffusion equation (depicted by the black dotted curve) at the time instant $t=2$   with parameters in equation \eqref{eq:parameters 1}. Moving to the right panel, we showcase the temporal evolution of the relative difference between the solution of the effective equation and the cross-sectional averaged stratified  scalar obtained through the full simulation. This relative difference is defined as the maximum deviation between the functions across the domain divided by the maximum value of the averaged stratified  scalar obtained from the full simulation.  }
  \label{fig:AveragedDensityComparisonFullSimulation}
\end{figure}

To proceed, we  simulate the effective equation \eqref{eq:effective equation}  with the same parameters \eqref{eq:parameters 1} for the comparison with the simulation of the governing equation \eqref{eq:NS nondimensional 3D diffusion driven flow rotated coordinate}.  In Figure \ref{fig:AveragedDensityComparisonFullSimulation}, the left panel offers a comprehensive comparison. Here, we superimpose the solution of the effective equation  with the cross-sectional averaged stratified scalar obtained through simulation of the complete governing equation \eqref{eq:NS nondimensional 3D diffusion driven flow rotated coordinate}  at the time instance $t=2$. The remarkable alignment between these curves serves to underscore the accuracy of the effective equation as an adept approximation for the full system. Additionally, we include the solution corresponding to the pure diffusion equation (representing the case in the absence of fluid flow) within the same visual representation. Notably, the contrast between the pure diffusion solution and the effective equation's solution demonstrates a visible enhancement of fluid mixing through diffusion-driven flow.

Shifting our attention to the right panel of Figure \ref{fig:AveragedDensityComparisonFullSimulation}, we delve into the temporal dynamics of the relative difference between the effective equation's solution and the full system's solution. At the inception ($t=0$), this disparity is nil, attributed to identical initial conditions. During the initial stages, the relative difference magnifies as the system has yet to transition into the regime well-captured by the effective equation. Nevertheless, even during this phase, the maximum relative difference remains at approximately 0.0056. As the system approaches an asymptotic state, which occurs after the diffusion time scale ($t=1$),  the relative difference diminishes, stabilizing at around 0.001 until the simulation's culmination. It's worth noting that due to the assumption of the asymptotic analysis, the effective equation is valid for small density gradient $\partial_{y_{1}}\bar{c} \ll 1$. In this numerical test case, we have $\max\limits_{y_{1},t} \partial_{y_{1}}\bar{c}= \frac{1}{\sqrt{\pi}}\approx 0.56419$, indicating that the parameter regime for the effective equation to reach a good approximation is larger than previously thought.

\begin{figure}
  \centering
    \includegraphics[width=1\linewidth]{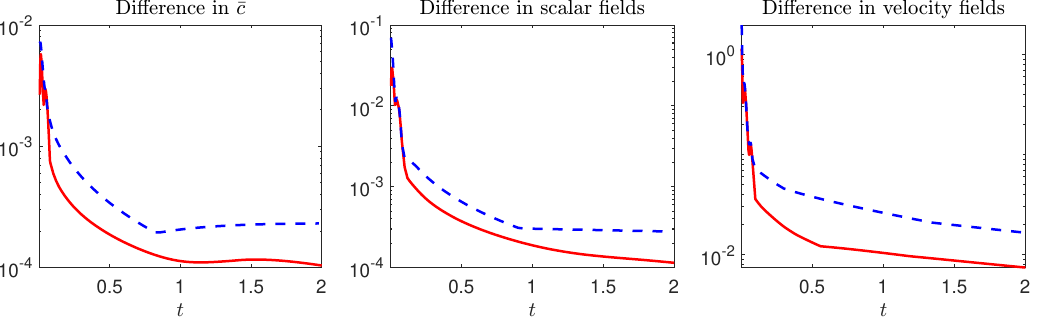}
  \hfill
  \caption[]
  {Comparison of the simulations with the initial condition in equation\eqref{eq:initial condition 1}, dimensionless parameter in \eqref{eq:parameters 1} and different $\mathrm{Sc}=1,10,100$. The red curve depicts the relative difference between the solution for $\mathrm{Sc}=1$ and the solution for $\mathrm{Sc}=10$. The blue curve depicts the relative difference between the solution for $\mathrm{Sc}=1$ and the solution for $\mathrm{Sc}=100$.}
  \label{fig:ComparisonFullSimulationDifferentSc}
\end{figure}

When deriving the effective equation in the previous section, we neglected the time derivative in equation \eqref{eq:diffusion-driven flow parallel-plate channel domain axillary problem1}. To verify the validity of this assumption, we conducted simulations with the same initial condition as in equation\eqref{eq:initial condition 1} and parameters in \eqref{eq:parameters 1}, but with different values of $\mathrm{Sc}=10,100$.   Figure \ref{fig:ComparisonFullSimulationDifferentSc} presents the relative differences in the scalar field, averaged scalar field, and velocity field  between the solution for $\mathrm{Sc}=10, 100$ and the solution for $\mathrm{Sc}=1$.   The relative differences for all three quantities are large at the early stage but quickly decreases as time increases.   At the diffusion time scale, the relative difference in scalar field is around $10^{-4}$ and the relative difference in velocity field is around $10^{-2}$, which approach the maximum accuracy achievable with the numerical scheme and resolution used in the simulation.   This observation suggest that the value of $\mathrm{Sc}$  does not have a significant impact on the scalar field and velocity field at the diffusion time scale, justifying the neglect of the time derivative in equation \eqref{eq:diffusion-driven flow parallel-plate channel domain axillary problem1} for $\mathrm{Sc}\geq 1$.

To demonstrate the validity of the effective equation across a wide range of parameters, we conducted a simulation with the same initial condition  \eqref{eq:initial condition 1}   and  a different set of nondimensional parameters,
\begin{equation}\label{eq:parameters 2}
\mathrm{Re}=0.1,\quad \mathrm{Ri}\mathrm{Re}=100000,\quad \mathrm{Pe}=100, \quad \mathrm{Sc}=1,\quad \theta=\frac{\pi}{4}.
\end{equation}
\begin{figure}
  \centering
  \subfigure[]{
    \includegraphics[width=0.46\linewidth]{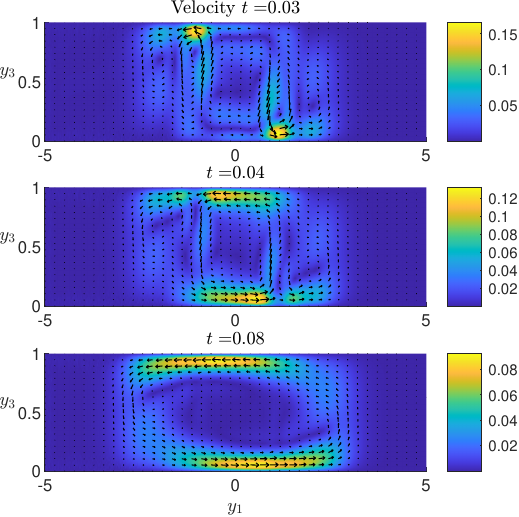}
  }
      \subfigure[]{
    \includegraphics[width=0.46\linewidth]{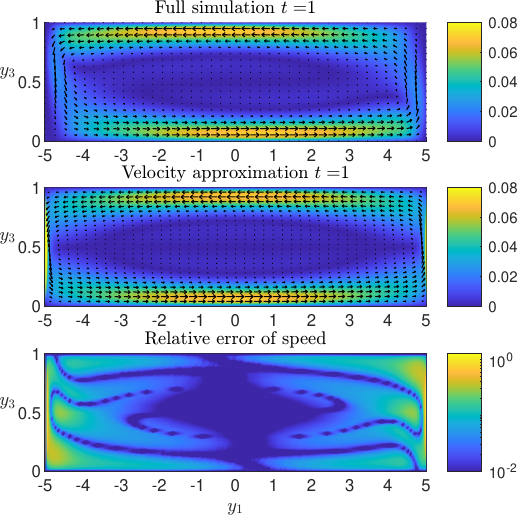}
  }
  \hfill
  \caption[]
  {In the left panel,  a pseudocolor plot illustrates the magnitude of the velocity field with parameters in equation \eqref{eq:parameters 2} at  three time instances: $t=0.03$ (top), $t=0.04$ (middle) and $t=0.08$ (bottom). In the right panel, the top figure displays the velocity field at $t=1$. The middle plot showcases the velocity field approximation at $t=1$, calculated using equations \eqref{eq:first order approximation density} and \eqref{eq:first order approximation v3}. Finally, the bottom plot depicts the relative approximation error of the speed.
  }
  \label{fig:FullSimulationtVelDensity1}
\end{figure}
The velocity field results are presented in panel (a) and at the top of panel (b) of  Figure \ref{fig:FullSimulationtVelDensity1}. Comparing Figure \ref{fig:FullSimulationtVelDensity} with Figure \ref{fig:FullSimulationtVelDensity1}, we make three observations: First, as $\mathrm{Ri}$ increases, the velocity field exhibits more intricate structures during the initial stages. A distinctive 'S' shape curve forms in the pseudocolor plot of the velocity field around $t=0.08$ and persists in subsequent time instants.  Second, the flow is more confined near the boundary, which is consistent with the observation we made from Figure \ref{fig:DiffusionDrivenFlowSteadySolution}. Third, at $t=1$, panel (b) of Figure \ref{fig:FullSimulationtVelDensity} illustrates that the velocity field is antisymmetric with respect to $y_{3}=\frac{1}{2}$. However, in panel (b) of Figure \ref{fig:FullSimulationtVelDensity1}, at $t=1$, the velocity field is not antisymmetric with respect to $y_{3}=\frac{1}{2}$ near the two ends  of the domain $y_{1}=\pm 5$.

The middle and bottom plots in Figure \ref{fig:FullSimulationtVelDensity1} (b) depict the approximation of the velocity field and the corresponding relative error at $t=1$. While the error remains relatively small within the interior of the domain, it increases notably near the corners and becomes significantly large in the narrow regions adjacent to the left and right boundaries, with the layer thickness approximately 0.05. This discrepancy stems from two main reasons. Firstly, the leading-order approximation of velocity exhibits antisymmetry across the entire domain, whereas the actual velocity field does not possess this property. Secondly, the asymptotic approximation outlined in the previous section holds true for an infinite domain. However, in a confined domain, the no-slip boundary condition for the velocity field at the end of the domain ($y_{3}=\pm 5$) is required but was not enforced during the derivation of the asymptotic approximation. While the no-flux boundary condition for the stratified scalar ensures a longitudinal velocity component of zero, the transverse velocity could deviate from zero where it should be, resulting in a significant error near the boundary. Achieving a more uniform approximation over the domain necessitates incorporating boundary layer corrections near the boundary into the velocity approximation.

\begin{figure}
  \centering
      \subfigure[]{
    \includegraphics[width=0.46\linewidth]{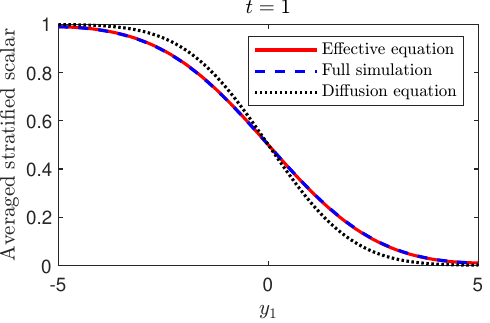}
  }
  \subfigure[]{
    \includegraphics[width=0.46\linewidth]{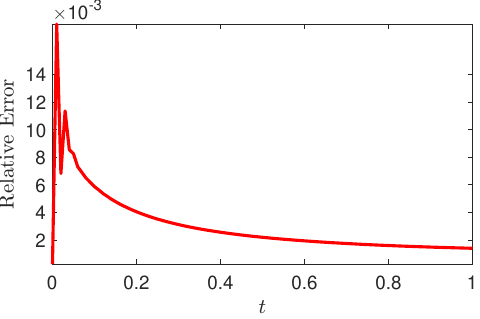}
  }
  \hfill
  \caption[]
  {In the left panel, we present the solution of the effective equation \eqref{eq:effective equation} (represented by the red solid curve)  with parameters in equation \eqref{eq:parameters 2}, the cross-sectional averaged stratified  scalar derived from simulating the complete governing equation (illustrated by the blue dashed curve), and the solution to the diffusion equation (depicted by the black dotted curve) at the time instant $t=1$. The right panel showcases the temporal evolution of the relative difference between the solution of the effective equation and the cross-sectional averaged stratified scalar obtained through the full simulation. }
  \label{fig:AveragedDensityComparisonFullSimulation1}
\end{figure}

One might assume that the substantial error in velocity approximation near the left and right ends of the domain could significantly compromise the accuracy of other approximations. However, as depicted in Figure \ref{fig:AveragedDensityComparisonFullSimulation1} (a), the solution of the effective equation perfectly overlaps with the averaged scalar field obtained from the full simulation. This alignment is further confirmed in Figure \ref{fig:AveragedDensityComparisonFullSimulation1} (b), illustrating the temporal dynamics of the relative difference, which remains small throughout the entire time interval.  We believe this is due to two possible reasons: Firstly, the error in the velocity field approximation is localized near the ends of the domain. Since the domain is long and narrow, the boundary layer constitutes a relatively small portion of the overall domain, consequently having a limited impact on the entire dynamic process. For a domain with comparable length scales in both directions, the error in the velocity field approximation may lead to a substantial error in the scalar field approximation.  Secondly, only the longitudinal component of the velocity contributes to the scalar transport in the leading-order approximation. Therefore, the error in the transverse component of velocity doesn't undermine the accuracy of the effective equation.

Lastly, we consider a staircase-like profile as the initial condition, a feature commonly encountered in certain oceanographic scenarios
\begin{equation}\label{eq:initial condition2}
\begin{aligned}
  &c_{I} =f (y_{1})+ \partial_{y_{1}}f(y_{1}) \cos \theta \left( y_{3}-\frac{1}{2} \right), \; f(y_{1})= \frac{ \mathrm{erfc}(y_{1}+\frac{5}{2})+\mathrm{erfc}(y_{1}-\frac{5}{2}) }{4},\; v_{1}=v_{3}=0. \\
\end{aligned}
\end{equation}
The nondimensional parameters are provided in equation \eqref{eq:non dimensional parameter1}  and $\theta=\frac{\pi}{4}$.

\begin{figure}
  \centering
  \subfigure[]{
    \includegraphics[width=0.46\linewidth]{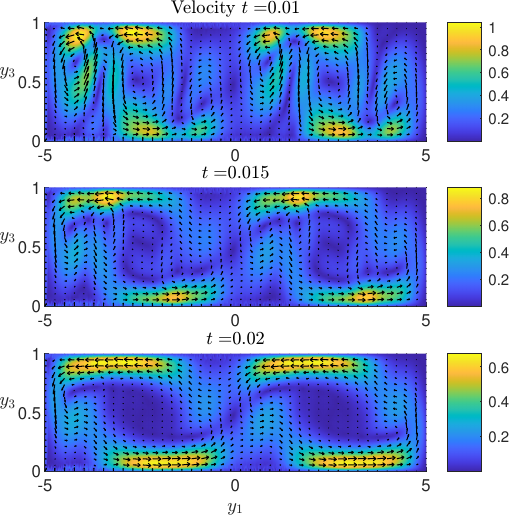}
  }
      \subfigure[]{
    \includegraphics[width=0.46\linewidth]{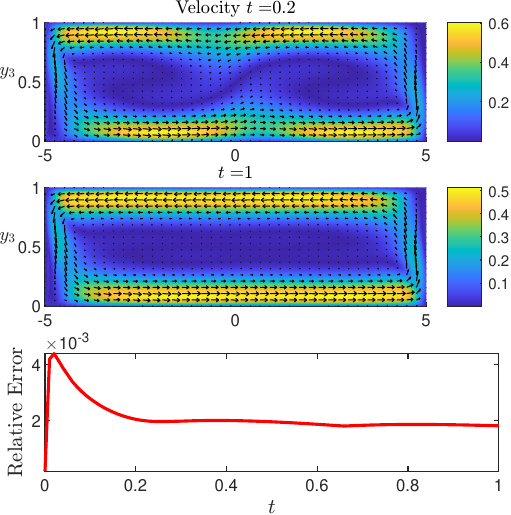}
  }
  \hfill
  \caption[]
  {A pseudocolor plot illustrates the magnitude of the velocity field with the initial condition \eqref{eq:initial condition2} and parameters in  \eqref{eq:non dimensional parameter1}  at different time instances. The bottom plot in panel (b)  showcases the temporal evolution of the relative difference between the solution of the effective equation and the cross-sectional averaged stratified scalar obtained through the full simulation.
  }
  \label{fig:FullSimulationtVelDensity2}
\end{figure}

The velocity field results are depicted in Figure \ref{fig:FullSimulationtVelDensity2}.  At the early phase ($t=0.01$), multiple convection cells emerges. The adjacent cells has differing orientations.  These cells swiftly amalgamate into two larger convection cells with the same anticlockwise orientation, located where the initial profile exhibits significant gradients. As the density distribution becomes smoother, around $t=0.2$, the merging of the two convection cells initiates. By $t=1$, only a single convection cell remains. The velocity distribution closely resembles that of simulations with differing initial conditions. The lower plot in Figure \ref{fig:FullSimulationtVelDensity2} (b) illustrates the disparity between the solutions of the effective equation and the cross-sectional averaged stratified scalar obtained through the full simulation.   The small relative difference again demonstrates the validity of the effective equation.

\section{ Analysis of the effective equation }
\label{sec:limit cases}
After confirming the accuracy of the effective equation \eqref{eq:effective equation} for the stratified scalar through numerical simulations, this section delves deeper into its properties.

\begin{figure}
  \centering
    \includegraphics[width=0.46\linewidth]{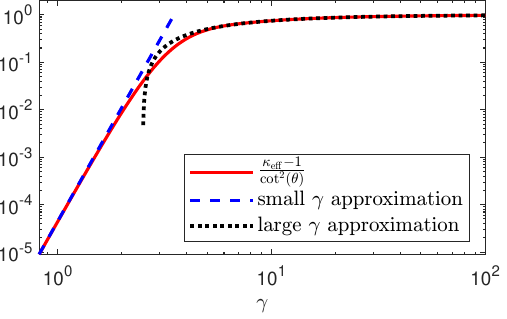}
  \hfill
  \caption[]
  { The red solid curve represents $\frac{\kappa_{\mathrm{eff}}-1}{\cot^{2} (\theta)}$ as a function of the nondimensional parameter $\gamma= \frac{ 1}{\sqrt{2}}\left( -\mathrm{Re} \mathrm{Pe} \mathrm{Ri}\sin \theta \partial_{y_{1}}\bar{\rho} \right)^{\frac{1}{4}}$. For small values of $\gamma$, the corresponding approximation is $\frac{\gamma ^8}{22680}$, which is depicted by the blue dashed curve. Conversely, for large values of $\gamma$, the corresponding approximation is $1- \frac{5}{2\gamma}$, illustrated by the black dotted curve. }
  \label{fig:diffusionDrivenKeGamma}
\end{figure}

The behavior of equation \eqref{eq:effective equation} is primarily governed by $\kappa_{\mathrm{eff}}$. Its asymptotic expansions for small and large values of $\gamma$ are provided below:
\begin{equation}
\begin{aligned}
  &\kappa_{\mathrm{eff}}=1 +\cot ^2(\theta ) \left(  \left(1-\frac{5}{2 \gamma }\right)+\frac{(2 \gamma +5) \sin (\gamma )+5 \cos (\gamma )}{\gamma  e^{\gamma }}+ \mathcal{O} \left( e^{-2\gamma}\right)\right) ,\; \gamma\rightarrow \infty,  \\
  &\kappa_{\mathrm{eff}}=1 +\cot ^2(\theta ) \left(  \frac{\gamma ^8}{22680}-\frac{2879 \gamma ^{12}}{4086482400}+ \mathcal{O} \left( \gamma^{16} \right) \right) ,\quad \gamma\rightarrow 0. \\
\end{aligned}
\end{equation}
Figure \ref{fig:diffusionDrivenKeGamma}  shows the graph  of $\frac{\kappa_{\mathrm{eff}}-1}{\cot^{2} (\theta)}$ and its approximations as a function of $\gamma$. By examining both the graph and the asymptotic approximations, we can conclude that for $\gamma\geq 0$: 

\begin{equation}
\begin{aligned}
1\leq \kappa_{\mathrm{eff}} (\gamma) < 1+ \cot^{2} (\theta).
\end{aligned}
\end{equation}

For large values of $\gamma$, equation \eqref{eq:effective equation} becomes  a diffusion equation with an enhanced diffusion coefficient:
\begin{equation}\label{eq:effective equation large gamma}
\begin{aligned}
&\partial_{t}\bar{c}=\left( 1+\cot^{2}\theta \right)  \partial_{y_{1}}^{2}\bar{c} .\\
\end{aligned}
\end{equation}
This resembles the scenario of a passive scalar governed by an advection-diffusion equation with a prescribed velocity field. In the context of the channel domain and at diffusion time scales, the advection-diffusion equation can be effectively simplified to a diffusion equation with an enhanced effective diffusion coefficient (see \cite{chatwin1970approach,smith1982contaminant,young1991shear,ding2021enhanced,ding2022ergodicity} for related work).

However, it's important to note that $\gamma$ is not a constant; it depends on the density gradient. Therefore, the diffusion equation \eqref{eq:effective equation large gamma} cannot provide a uniform approximation of the effective equation for all time instances. As time progresses, the stratified scalar becomes more homogeneous across the domain. As the density gradient decreases, $\gamma$ decreases, and the approximation given by \eqref{eq:effective equation large gamma} becomes less accurate.

\begin{figure}
  \centering
    \includegraphics[width=1\linewidth]{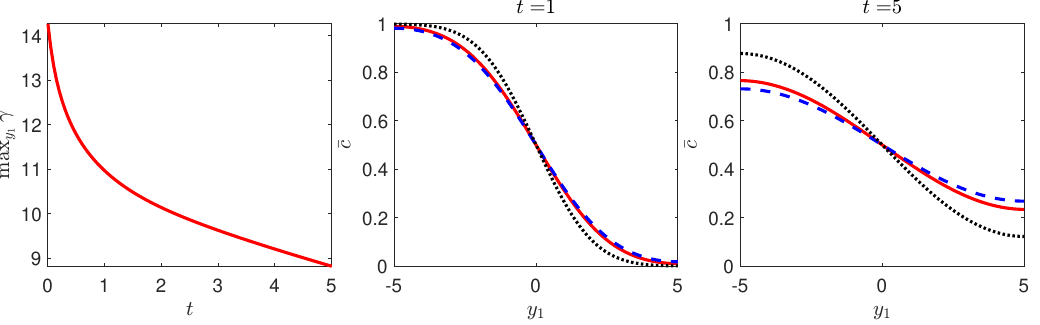}
  \hfill
  \caption[]
  {  The left panel shows $\max\limits_{y_{1}}\gamma$ as a function of time in the simulation with the parameters provided in equation \eqref{eq:parameters 2}. The middle panel  presents the solution of the effective equation \eqref{eq:effective equation} (represented by the red solid curve), the solution of diffusion equation with diffusivity $1+\cot^{2} (\theta)$  (illustrated by the blue dashed curve), and the solution to the diffusion equation with diffusivity 1 (depicted by the black dotted curve) at the time instant $t=1$. The right panel shows the corresponding curves at the time instant $t=5$. }
  \label{fig:CompareDiffusion}
\end{figure}

To illustrate this, we conducted simulations with the parameters provided in equation \eqref{eq:parameters 2} as an example. The left panel in figure \ref{fig:CompareDiffusion} shows the maximum value of $\gamma$ across the domain as a function of time. As we expect, it is a decreasing function. The middle panel of figure \ref{fig:CompareDiffusion} demonstrates that the solution to the diffusion equation \eqref{eq:effective equation large gamma} reasonably approximates the solution to the effective equation \eqref{eq:effective equation} initially. However, as time progresses and the density gradient decreases, $\max\limits_{y_{1}}\gamma$ also decreases. At $t=5$, the difference between the solution of equation \eqref{eq:effective equation large gamma} and that of equation \eqref{eq:effective equation} becomes more pronounced. This example also demonstrate that a single diffusion equation is not enough to accurately describes the dynamics of the averaged stratified  scalar at all different time scale.

Equation \eqref{eq:effective equation large gamma} suggests that as the inclination angle approaches zero, the dispersion rate may increase, possibly reaching infinity. It's crucial to emphasize that this conclusion holds true only when other parameters are kept constant.  It's worth noting that as the inclination angle decreases, maintaining the same value of $\gamma$ necessitates keeping the same density variation in the $y_{1}$ direction, which in turn requires a larger density variation in the $x_{3}$ direction. Achieving such a condition in practical applications can pose significant challenges.

Next we consider the case with small $\gamma$.   In this limit,  the effective equation for the stratified scalar \eqref{eq:effective equation} reduces to
\begin{equation}\label{eq:effective equation small density gradient}
\begin{aligned}
  &\partial_{t}\bar{c}=\partial_{y_{1}} \left( \left( 1+\left( \frac{ \mathrm{Ri}\text{Pe} \text{Re} \cos (\theta )}{72 \sqrt{70} }    \partial_{y_{1}} \rho(\bar{c}) \right)^{2}  \right) \partial_{y_{1}}\bar{c} \right). \\
\end{aligned}
\end{equation}
The closed form expression of the solution of this equation is not available. However, we can obtain the solution in some limits for unbounded domain. When the nonlinear term is much larger than the diffusion term, namely, $\frac{ \mathrm{Ri}\text{Pe} \text{Re} \cos (\theta )}{72 \sqrt{70} }    \partial_{y_{1}}\rho(\bar{c})  \gg 1$, and $\partial_{c}\rho$ is a constant,  equation \eqref{eq:effective equation small density gradient} can be approximated by
\begin{equation}\label{eq:effective equation small density gradient 1}
\partial_{t}\bar{c}=\partial_{y_{1}} \left( \alpha^{2}  \left( \partial_{y_{1}}\bar{c} \right) ^{3} \right), \quad \alpha=\frac{ \mathrm{Ri}\text{Pe} \text{Re} \cos (\theta ) \partial_{c}\rho}{72 \sqrt{70} }. 
\end{equation}
 We can look for the self similarity solution, $c (y_{1},t)=f \left( \xi \right)$, where $\xi=t y_{1}^{-\frac{1}{4}}$. Equation \eqref{eq:effective equation small density gradient 1} becomes
\begin{equation}
  f'(\xi ) \left(12 \alpha^{2}  f'(\xi ) f''(\xi )+\xi \right)=0.
\end{equation}
The solution is
\begin{equation}
\begin{aligned}
&f (\xi)=
\begin{cases}
  C_{2}+ \pi  C_{1} \alpha \sqrt{3} & \xi \geq 2 \alpha \sqrt{6  C_{1}},\\
 C_{2}- \frac{\frac{1}{2} \xi  \sqrt{24 \alpha^{2}  c_1-\xi ^2}+12 \alpha^{2}  c_1 \tan ^{-1}\left(\frac{\xi }{\sqrt{24 \alpha^{2}  c_1-\xi ^2}}\right)}{2 \sqrt{3} \sqrt{\alpha }}&- 2 \alpha \sqrt{6  C_{1}} \leq \xi \leq 2 \alpha \sqrt{6  C_{1}} ,\\
  C_{2}- \pi  c_1 \alpha \sqrt{3}& \xi \leq -2 \alpha \sqrt{6  C_{1}},
\end{cases}
\end{aligned}
\end{equation}
where $C_{1}$, $C_{2}$ are the constants that can be defined by the boundary condition at infinity. For the case $\bar{c} (\infty)=0$ and $\bar{c} (-\infty)=1$, we have $C_{1}=\frac{1}{2 \sqrt{3} \pi  \alpha }$,  $C_{2}=\frac{1}{2}$. Notice that the similarity variable for the pure diffusion equation is $\xi=\frac{y_{1}}{\sqrt{t}}$. Therefore, $t\rightarrow \infty$, the neglected diffusion term in the above calculation becomes dominant eventually.

Last, we can represent $\gamma$ in terms of the physical parameters as follows:
$\gamma = \frac{1}{\sqrt{2}} \left( \frac{g H^3 \rho_{0}}{\kappa \mu} \sin \theta \partial_{y_{1}}\rho (\bar {c}) \right)^{\frac{1}{4}}$. Therefore, smaller diffusivity, lower viscosity, greater gravitational constant, larger gap thickness, and higher density result in a larger value of $\gamma$.  We can calculate $\gamma$ using the practical parameters provided in equation \eqref{eq:non dimensional parameter1}. If we set $\theta=\frac{\pi}{4}$ and estimate the density gradient as $ \partial_{y_{1}}\rho (\bar{c}) =\sin \theta \partial_{x_{3}} \rho (\bar{c})$, we obtain $\gamma \approx 8.94506$ and $\kappa_{\mathrm{eff}} (\gamma) \approx1.720605$.  With the same parameters and a smaller inclination angle $\theta=\frac{\pi}{10}$, we have $\gamma \approx 5.91333$ and a larger effective diffusivity $\kappa_{\mathrm{eff}} (\gamma) \approx 6.46129$. In these cases, we cannot employ the approximate equations \eqref{eq:effective equation large gamma} and \eqref{eq:effective equation small density gradient}. Instead, we must use the complete effective equation \eqref{eq:effective equation} to solve for the stratified scalar.

\subsection{Long time behavior of the solution}
As demonstrated in our previous examples, diffusion-driven flow enhances the mixing of a stratified scalar at some time scales.  One might expect the difference between the density profile with and without the fluid to increase as time progresses or at least persist at long times. However, intriguingly, this difference actually vanishes over extended periods.

To understand this, it's essential to recall an interesting observation that emerges when considering different solutions to the same diffusion equation. Under some conditions, these solutions  converge to a self-similarity solution asymptotically at long times, as discussed in  \cite{newman1984lyapunov}.  To illustrate this point, consider two solutions: $f_{1}=\frac{1}{2} \text{erfc}\left(\frac{x}{2 \sqrt{t}}\right)$ and $f_{2}=\frac{1}{2} \text{erfc}\left(\frac{x}{2 \sqrt{\sigma ^2+t}}\right)$, both satisfying the diffusion equation $\partial_{t}f=\partial_{x}^{2}f$. Although they begin with different initial conditions, the relative difference between them diminishes as $t\rightarrow \infty$:
\begin{equation}
\begin{aligned}
\frac{f_{1}-f_{2}}{f_{1}}= \frac{\sigma ^2 x t^{-\frac{3}{2}} }{2 \sqrt{\pi }}+ \mathcal{O} \left( t^{-2} \right).
\end{aligned}
\end{equation}
The timescale for this convergence depends on the difference in the initial conditions and can be multiple times the diffusion time scale.

This observation implies that even if diffusion-driven flow initially amplifies the dispersion of the stratified agent, when the density gradient is sufficiently weak, the governing equation for the stratified scalar approximates a diffusion equation with a diffusivity of 1. While some disparity remains between the solution of the effective equation and the solution of the pure diffusion equation, this difference diminishes over longer time scales.

\begin{figure}
  \centering
  \subfigure[]{
    \includegraphics[width=0.46\linewidth]{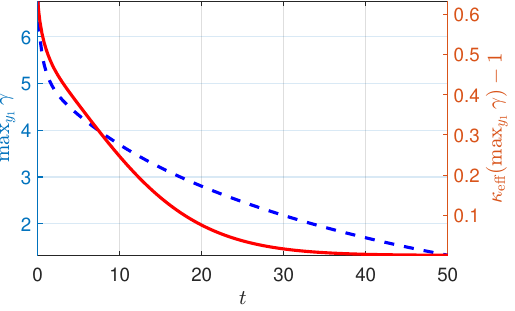}  }
  \subfigure[]{
    \includegraphics[width=0.46\linewidth]{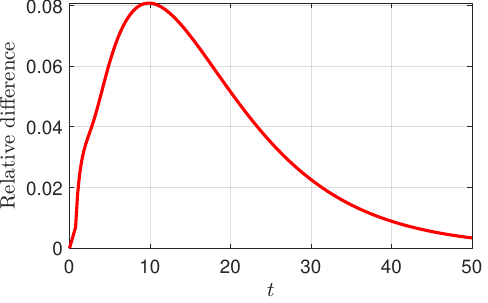}
  }
  \hfill
  \caption[]
  {(a)The blue dashed curve represents $\max_{y_{1}}\gamma $ and is associated with the left $y$ axis. The red solid curve represents $\kappa_{\mathrm{eff}} (\max_{y_{1}}\gamma )-1$ and is associated with the right $y$ axis. (b) The relative difference between the density profile with and without the fluid as a function of time. The parameters used in the simulation are provided in equation \eqref{eq:parameters 1}.}
  \label{fig:ConvergeToDiffusion}
\end{figure}

To confirm this conclusion, we solve the effective equation \eqref{eq:effective equation} with parameters provided in equation \eqref{eq:parameters 1}. In Panel (a) of Figure \ref{fig:ConvergeToDiffusion}, we plot $\max_{y_{1}}\gamma$ and $\kappa_{\mathrm{eff}} (\max_{y_{1}}\gamma )-1$ as functions of time, providing an estimation of the enhancement of dispersion induced by diffusion-driven flow. We observe that the enhanced diffusivity decreases below 0.1 after $t>20$, which implies the effective equation is close to the pure diffusion equation after that time.  In Panel (b) of Figure \ref{fig:ConvergeToDiffusion}, we display the relative difference between the density profiles with and without flow. This difference initially increases, reaching its maximum around $t=10$, but subsequently decreases to 0.003 at $t=50$.

This result demonstrate that in a confined domain with no-flux boundary condition, the density profile in a system with the diffusion driven flow asymptotically converges to the one without the flow at extremely large time scale.

\section{Comparison to the thin film approximation}
\label{sec:power series expansion}
As previously mentioned, we can employ the classical asymptotic expansion to analyze the governing equation \eqref{eq:NS nondimensional 3D diffusion driven flow rotated coordinate}. In this section, we will utilize the thin film approximation and subsequently compare the results with those obtained previously.

In the thin film approximation, we select distinct length scales in different directions for the purpose of nondimensionalization:
\begin{equation}
\begin{aligned}
  &L y_{1}'=y_{1}, \quad H y_{3}'=y_{3},  \quad  V_{i} v_{i}'=v_{i}, \quad  \frac{\rho_{0}L\nu U}{H^{2}}p'=p,  \quad \frac{L^{2}}{\kappa} t'=t, \quad \rho_{0}\rho'=\rho, \\
  &c_{0}c'=c,\quad \epsilon=\frac{V_{3}}{V_{1}}=\frac{H}{L} \ll 1, \quad \mathrm{Re}=\frac{ V_{1} H}{\nu},\quad \mathrm{Pe}= \frac{V_{1}H}{\kappa}, \quad \mathrm{Sc}= \frac{\mu}{\rho_{0} \kappa},\quad \mathrm{Ri}= \frac{ H g}{V_{1}^{2}}.
\end{aligned}
\end{equation}
The nondimensionalized governing equation becomes
\begin{equation}
\begin{aligned}
&\rho' \left( \frac{\epsilon^{2}}{\mathrm{Sc}}\partial_{t'}  v_{1}'+\epsilon \mathrm{Re} \left(    v_{1}'\partial_{y_{1}'} v_{1}' +  v_{3}'\partial_{y_{3}'} v_{1}'\right) \right) = \left( \epsilon^{2}\partial_{y_{1}'}^{2} v_{1}' + \partial_{y_{3}'}^{2} v_{1}'  \right)- \partial_{y_{1}'} p'-  \mathrm{Re}  \mathrm{Ri}\rho'  \sin \theta,  \\
&\rho'\left( \frac{\epsilon^{3}}{\mathrm{Sc}}\partial_{t'}  v_{3}'+ \epsilon^{2} \mathrm{Re}\left(v_{1}'\partial_{y_{1}'} v_{3}' +  v_{3}'\partial_{y_{3}'} v_{3}'\right)  \right)= \left(  \epsilon^{3} \partial_{y_{1}'}^{2} v_{3}' + \epsilon \partial_{y_{3}'}^{2} v_{3}' \right) - \frac{1}{\epsilon}\partial_{y_{3}'} p'-  \mathrm{Re} \mathrm{Ri} \rho'  \cos \theta,  \\
&\epsilon^{2}\partial_{t'}c'+ \epsilon \mathrm{Pe} v_{1}'\partial_{y_{1}'}c'+ \epsilon\mathrm{Pe} v_{3}'\partial_{y_{3}'}c'= \epsilon^{2} \partial_{y_{1}'}^{2}c'+  \partial_{y_{3}'}^{2}c',\\
&\partial_{y_{1}'}v_{1}'+\partial_{y_{3}'}v_{3}'=0.\\
\end{aligned}
\end{equation}
Dropping primes and rearranging the equation results
\begin{equation}
  \begin{aligned}
&\rho\left( \frac{\epsilon^{2}}{\mathrm{Sc}}\partial_{t}  v_{1}+\epsilon \mathrm{Re} \left(    v_{1}\partial_{y_{1}} v_{1} +  v_{3}\partial_{y_{3}} v_{1}\right)  \right)= \left( \epsilon^{2}\partial_{y_{1}}^{2} v_{1} + \partial_{y_{3}}^{2} v_{1}  \right)- \partial_{y_{1}} p-  \mathrm{Re} \mathrm{Ri} \rho  \sin \theta,  \\
&\rho \left( \frac{\epsilon^{4}}{\mathrm{Sc}}\partial_{t}  v_{3}+ \epsilon^{3} \mathrm{Re}\left(  v_{1}\partial_{y_{1}} v_{3} +  v_{3}\partial_{y_{3}} v_{3}\right)  \right)= \epsilon^{2}\left(  \epsilon^{2} \partial_{y_{1}}^{2} v_{3} +  \partial_{y_{3}}^{2} v_{3} \right) - \partial_{y_{3}} p- \epsilon \mathrm{Re} \mathrm{Ri} \rho  \cos \theta,  \\
&\epsilon^{2}\partial_{t}c+\epsilon \mathrm{Pe} \left(  v_{1}\partial_{y_{1}}c+   v_{3}\partial_{y_{3}}c \right)= \epsilon^{2} \partial_{y_{1}}^{2}c+  \partial_{y_{3}}^{2}c,\\
&\partial_{y_{1}}v_{1}+\partial_{y_{3}}v_{3}=0.\\
\end{aligned}
\end{equation}
We consider the formal power series expansions for the velocity, stratified scalar and pressure, \begin{equation}
\begin{aligned}
  &v_{i}=\sum\limits_{k=0}^{\infty}v_{i,k}\epsilon^{k}, \quad c=\sum\limits_{k=0}^{\infty}c_{k}\epsilon^{k},\quad  p=\sum\limits_{k=0}^{\infty}p_{k}\epsilon^{k}, \quad \rho (c)= \sum\limits_{k=0}^{\infty} \rho_{k}\epsilon^{k}.
\end{aligned}
\end{equation}
We have $\left. v_{i,k} \right|_{\mathbf{y}\in \partial\Omega }=0$ from the no-slip boundary condition of the velocity field, and $\left. \partial_{\mathbf{n}} c_{k}\right|_{\mathbf{y} \in \partial \Omega}=0$ from the no-flux boundary condition of the stratified scalar. The coefficient in the expansion of $\rho$ can be calculated from $c_k$ and the Taylor expansion of $\rho (c)$.

Substituting the expansion into the governing equation and collecting $\mathcal{O}(1)$ terms yield the following equation
\begin{equation}
\begin{aligned}
&0= \partial_{y_{3}}^{2} v_{1,0}  - \partial_{y_{1}} p_{0}-  \mathrm{Re} \mathrm{Ri} \rho_{0}  \sin \theta, \; 0=\partial_{y_{3}}p_{0},\; 0 = \partial_{y_{3}}^{2}c_{0},\; \partial_{y_{1}}v_{1,0}+\partial_{y_{3}}v_{3,0}=0.
\end{aligned}
\end{equation}
There are two possible group of solutions:
\begin{equation}\label{eq:O0 solution}
\begin{aligned}
&v_{1,0}=v_{3,0}=0, \quad p_{0}=  \mathrm{Re} \mathrm{Ri}   \sin \theta  \int\limits_0^{y_{1}}\rho_{0} \mathrm{d} y_{1},
\end{aligned}
\end{equation}
and
\begin{equation}\label{eq:O0 solution pressure driven flow}
\begin{aligned}
  &v_{1,0}=C y_{3} (y_{3}-1), \;  v_{3,0}=0, \quad p_{0}= -2Cy_{1}+ \mathrm{Re} \mathrm{Ri}   \sin \theta  \int\limits_0^{y_{1}}\rho_{0} \mathrm{d} y_{1},
\end{aligned}
\end{equation}
where $C$ is a constant number and $c_{0}, \rho_0$ are independent of $y_{3}$. The second solution describes a flow generated by a constant pressure gradient, which does not represent the correct physical problem. Therefore, we regard the first solution as the correct leading-order approximation.

Collecting the terms that is comparable to $\mathcal{O}(\epsilon)$ yields the following equation
\begin{equation}
\begin{aligned}
& \mathrm{Re}\rho_{0} \left(    v_{1,0}\partial_{y_{1}} v_{1,0}+  v_{3,0}\partial_{y_{3}} v_{1,0}\right) = \partial_{y_{3}}^{2} v_{1,1} - \partial_{y_{1}} p_{1}- \mathrm{Re} \mathrm{Ri} \rho_{1}  \sin \theta,  \\
&0=-\partial_{y_{3}} p_{1}-\mathrm{Re} \mathrm{Ri} \rho_{0}  \cos \theta,  \quad  \mathrm{Pe} \left(  v_{1,0}\partial_{y_{1}}c_{0} +  v_{3,0}\partial_{y_{3}}c_{0}  \right)=  \partial_{y_{3}}^{2}c_{1},\\
&\partial_{y_{1}}v_{1,1}+\partial_{y_{3}}v_{3,1}=0.
\end{aligned}
\end{equation}
Using the conclusions \eqref{eq:O0 solution} reduces the above equation to
\begin{equation}\label{eq:O1 equation2}
\begin{aligned}
  &0 = \partial_{y_{3}}^{2} v_{1,1} - \partial_{y_{1}} p_{1}-  \mathrm{Re}\mathrm{Ri} \rho_{1}  \sin \theta,  \quad 0=-\partial_{y_{3}} p_{1}-\mathrm{Re} \mathrm{Ri} \rho_{0}  \cos \theta, \\
  &  0=  \partial_{y_{3}}^{2}c_{1},\quad \partial_{y_{1}}v_{1,1}+\partial_{y_{3}}v_{3,1}=0.
\end{aligned}
\end{equation}
 Similar as the equation for $\mathcal{O} (1)$ terms, the solution is not unique. To eliminate the solution representing the pressure driven flow, we impose the condition $\bar{p}_{1}=0$.  From the second equation, we have
\begin{equation}
\begin{aligned}
&p_{1}= -\left( y_{3}-\frac{1}{2} \right)\mathrm{Re} \mathrm{Ri} \rho_{0}  \cos \theta.
\end{aligned}
\end{equation}
Since $c_{1}$ is independent of $y_{3}$, it follows that $\rho_{1}=c_{1}\partial_{c}\rho (c_{0})$ is also independent of $y_{3}$.
Averaging the first equation yields $\bar{v}_{1,1}=0$. Differentiating the first equation in equation \eqref{eq:O1 equation2} with respect to $y_{3}$ twice give us
\begin{equation}
  \begin{aligned}
  &0 = \partial_{y_{3}}^{4} v_{1,1}, \quad \left. v_{1,1} \right|_{y=0,1 }=0,\quad  \left. \partial_{y_{3}}^{3}v_{1,1} \right|_{y=0,1 }=- \mathrm{Re}\mathrm{Ri}  \cos \theta \partial_{y_{1}}  \rho_{0} . 
\end{aligned}
\end{equation}
The solution is 
\begin{equation}
\begin{aligned}
&v_{1,1}=- \mathrm{Re}\mathrm{Ri} \cos \theta \partial_{y_{1}}  \rho_{0}   \frac{y_{3}  (y_{3}-1) (2 y_{3}-1)}{12},
\end{aligned}
\end{equation}
which is consistent with the first term in equation \eqref{eq:velocity low density gradient}. It's important to note that the definition of $v_{1,1}$ in equation \eqref{eq:velocity low density gradient} differs from the one presented here. To make a comparison, $v_{1,1}$ in \eqref{eq:velocity low density gradient} need to be multiplied with $\partial_{y_{1}}\bar{c}$. 

From the third equation in \eqref{eq:O1 equation2}, we conclude that $c_{1}$ is independent of $y_{3}$. Moreover, after substituting the expression of $v_{1,1}$ back to the first equation in equation \eqref{eq:O1 equation2}, we obtain $\rho_{1}=0$. The continuity equation gives us the expression of $v_{3,1}$:
\begin{equation}
\begin{aligned}
&v_{3,1}=-\int\limits_0^{y_{3}} \partial_{y_{1}}v_{1,1}\mathrm{d} y_{3}=  \mathrm{Re}\mathrm{Ri} \cos \theta \partial_{y_{1}}^{2}  \rho_{0}   \frac{y_{3} ^{2} (y_{3}-1) ^{2}}{24},\\
\end{aligned}
\end{equation}
which is consistent with the first term in equation \eqref{eq:first order approximation v3}.

Collecting the terms that are comparable to $\mathcal{O}(\epsilon^{2})$ yields the following equation
\begin{equation}
  \begin{aligned}
    &\frac{\rho_{0}}{\mathrm{Sc}}\partial_{t}  v_{1,0}+\mathrm{Re} \rho_{0}\left(   v_{1,0}\partial_{y_{1}} v_{1,1}+ v_{1,1}\partial_{y_{1}} v_{1,0} +  v_{3,0}\partial_{y_{3}} v_{1,1}+  v_{3,1}\partial_{y_{3}} v_{1,0}\right) \\
    &+ \mathrm{Re} \rho_{1}\left(   v_{1,0}\partial_{y_{1}} v_{1,0} +  v_{3,0}\partial_{y_{3}} v_{1,0}  \right)= \left(\partial_{y_{1}}^{2} v_{1,0} + \partial_{y_{3}}^{2} v_{1,2}  \right)- \partial_{y_{1}} p_{2}-  \mathrm{Re}\mathrm{Ri} \rho_{2}  \sin \theta,  \\
&0 =  \partial_{y_{3}}^{2} v_{3,0} - \partial_{y_{3}} p_{2}-  \mathrm{Re}\mathrm{Ri} \rho_{1}  \cos \theta,  \\
&\partial_{t}c_{0}+ \mathrm{Pe} \left(   v_{1,0}\partial_{y_{1}}c_{1}+  v_{1,1}\partial_{y_{1}}c_{0}+   v_{3,0}\partial_{y_{3}}c_{1} + v_{3,1}\partial_{y_{3}}c_{0}  \right)=  \partial_{y_{1}}^{2}c_{0}+  \partial_{y_{3}}^{2}c_{2},\\
&\partial_{y_{1}}v_{1,2}+\partial_{y_{3}}v_{3,2}=0.
\end{aligned}
\end{equation}

Using the current conclusions for $\mathcal{O} (\epsilon)$ terms, the above equation reduces to
\begin{equation}\label{eq:O2 equation2}
  \begin{aligned}
&0 = \partial_{y_{3}}^{2} v_{1,2}  -\partial_{y_{1}} p_{2}-  \mathrm{Re}\mathrm{Ri} \rho_{2}  \sin \theta,  \quad 0 =   - \partial_{y_{3}} p_{2}-  \mathrm{Re}\mathrm{Ri} \rho_{1}  \cos \theta,  \\
&\partial_{t}c_{0}+ \mathrm{Pe} v_{1,1}\partial_{y_{1}}c_{0}=  \partial_{y_{1}}^{2}c_{0}+  \partial_{y_{3}}^{2}c_{2}.\\
\end{aligned}
\end{equation}
Notice that  $\int\limits_0^{1}v_{1,1} \mathrm{d} y_{3}=0$ and $c_0$ is independent of $y_{3}$. Integrating on both side of the last equation in equation \eqref{eq:O2 equation2} with respect to $y_{3}$ and using the no-flux boundary conditions  yields  the evolution equation for $c_{0}$: $\partial_{t}c_{0}=  \partial_{y_{1}}^{2}c_{0}$.

Now the last equation in equation \eqref{eq:O2 equation2} becomes $ \mathrm{Pe} v_{1,1}\partial_{y_{1}}c_{0}= \partial_{y_{3}}^{2}c_{2}$ and implies
\begin{equation}
\begin{aligned}
&c_{2}=\frac{\mathrm{Pe}\text{Re} \text{Ri} \left(1-2  y_{3}^3(10+3 y_{3} (2 y_{3}-5))\right) \cos (\theta ) \partial_{y_{1}}c_{0} \partial_{y_{1}} \rho_{0} }{1440}.
\end{aligned}
\end{equation}
which is consistent with the second term in equation \eqref{eq:rho low density gradient}. It's important to note that the definition of $c_{1}$ in section \ref{sec:Flow for slowly varying density profile} differs from the one presented here.

The stratified scalar approximation is expressed as $c=c_{0}+\epsilon^{2}c_{2}+\mathcal{O} (\epsilon^{3})$, which serves as the solution to a pure diffusion equation augmented by a higher order correction term. The fact that $\bar{c}_{2}=0$ implies that as $\epsilon$ tends to zero, the diffusion-driven flow only distorts the contour line of the stratified scalar without amplifying the dispersion of the stratified  scalar in the longitudinal direction of the channel.

So far, we have observed that the results obtained from the thin film approximation are consistent with the results obtained in Section \ref{sec:Flow for slowly varying density profile} when the density gradient vanishes, i.e., $\partial_{y_{1}}\bar{\rho}\rightarrow 0$, or equivalently when $\gamma\rightarrow 0$. Therefore, the thin film approximation presented here may not accurately model systems where the diffusion-driven flow significantly enhances dispersion, as is the case with the parameters provided in Equation \eqref{eq:parameters 2}, where the corresponding $\gamma =9\sim 15$ and the flow make visible enhancement of the scalar dispersion as shown in figure \ref{fig:CompareDiffusion}. For those parameter combinations, an asymptotic analysis using a  scaling relation that differs from the one presented in this section become necessary.

\section{Conclusion and discussion}
\label{sec:conclusion}
This paper explores the diffusion-driven flow in a tilted parallel-plate channel domain with a nonlinear density stratification. By employing a novel asymptotic expansion provided in equation \eqref{eq:asymptotic expansion ansatz}, we derive leading-order approximations for the velocity field \eqref{eq:first order approximation density} and \eqref{eq:first order approximation v3}, stratified scalar \eqref{eq:first order approximation density}, and pressure field \eqref{eq:first order approximation pressure}. Furthermore, we formulate an effective equation \eqref{eq:effective equation} to describe the cross-sectional averaged stratified scalar, and its accuracy is confirmed through numerical simulations of the governing equation \eqref{eq:NS nondimensional 2D diffusion driven flow Boussinesq}.

The effective equation reveals that the dynamics of the stratified scalar depend on the dimensionless parameter $\gamma$, defined as $\gamma= \frac{1}{\sqrt{2}}\left(-\mathrm{Re} \mathrm{Pe} \mathrm{Ri} \sin \theta \partial_{y_{1}}\rho (\bar{c})\right)^{\frac{1}{4}}= \frac{1}{\sqrt{2}} \left(\frac{g L^3 \rho_{0}}{\kappa \mu} \sin \theta \partial_{y_{1}} \rho (\bar{c})\right)^{\frac{1}{4}}$. When $\gamma$ is large, the system behaves akin to a diffusion equation with an enhanced diffusion coefficient of $1+\cot^{2}\theta$, where $\theta$ is the  inclination angle  relative to the horizontal plane. This result reveals an upper bound for the mixing capability of  the diffusion-driven flow. Conversely, in the small $\gamma$ limit, the behavior of the stratified scalar approximates a pure diffusion process, with the diffusion-driven flow failing to amplify the dispersion of the stratified scalar significantly. Additionally, we demonstrate that in a confined domain with no-flux boundary conditions, the density profile in a system featuring diffusion-driven flow asymptotically converges to the one without flow over long times, although this convergence occurs on a timescale much larger than the diffusion time scale.

Moreover, we establish that the thin film approximation aligns with the results obtained using the novel expansion when the diffusion-driven flow is weak. In such scenarios, the stratified scalar can be modeled by a diffusion equation featuring molecular diffusivity, and the diffusion-driven flow primarily distorts the scalar distribution without significantly increasing longitudinal dispersion. Consequently, the thin film approximation falls short in describing systems with relatively large density gradients, where diffusion-driven flow markedly enhances dispersion. Importantly, we both numerically and theoretically demonstrate that the proposed expansion effectively addresses these situations.

Future research directions encompass several avenues.  Firstly, the passive scalar transport in the channel with non-flat boundaries has various practical applications (\cite{roggeveen2023transport,chang2023taylor}).  Therefore, our aim is to develop a reduced model of diffusion-driven flow in channels with rough boundaries, such as rock fissures or microfluidic devices. When the amplitude and wavelength of the boundary variation are small, the rough boundary with a no-slip boundary condition can be approximated by a flat wall with an effective slip boundary condition obtained through the multiscale method (\cite{achdou1998effective,carney2022heterogeneous}). In this scenario, the method presented in our work can be readily applied. Alternatively, when the wavelength of the boundary variation is large, the asymptotic expansion presented by \cite{mercer1990centre}  could offer a solution. Secondly, while this work primarily focuses on analyzing parallel plate domains, it's worth noting that the method proposed herein is applicable to channels with arbitrary cross-sections. Concentrating on the parallel plates domain stems from the availability of an exact solution for the auxiliary problem. Investigating this fundamental geometry can augment our understanding of domains with more complex shapes, such as tilted cylindrical cavities embedded in rocks \cite{sanchez2005natural}, or tilted square container (\cite{grayer2020dynamics,page2011combined,page2011steady,french2017diffusion}). 

\section{Acknowledgements}
I would like to thank Professor A.J. Roberts for his insightful comments on the application of the slow manifold theory (\cite{roberts1988application,roberts2014model,roberts2015macroscale}) to the problem discussed in this paper. I also extend my gratitude to the anonymous referees, whose feedback has significantly improved the quality of this paper.

\section{ Declaration of Interests}
The authors report no conflict of interest.

\appendix
\section{Numerical method}\label{sec:appendix numerical method}
In this section, we document the numerical method used for solving the Navier-Stokes equation \eqref{eq:NS nondimensional 3D diffusion driven flow rotated coordinate} and the effective equation \eqref{eq:effective equation} .

To solve the Navier-Stokes equation \eqref{eq:NS nondimensional 3D diffusion driven flow rotated coordinate}, we employ the projection algorithm. During each iteration, we explicitly evaluate the advection term, while treating the viscosity term implicitly. This involves solving a Poisson equation while enforcing the no-slip boundary condition. Subsequently, we solve the pressure Poisson equation, wherein adjustments are made to the velocity field to ensure fulfillment of the incompressibility condition. A comprehensive outline of the numerical scheme can be found in \cite{hecht2005freefem++}. We employ a similar approach for the advection-diffusion equation. In each step, the advection term is explicitly computed, whereas the diffusion term is treated implicitly, necessitating the resolution of a Poisson equation with a no-flux boundary condition. The time-stepping scheme used in this algorithm results in first-order accuracy in time. The finite element method is used to discretize the system of equations, and we implement the algorithm using the software FreeFEM++ (\cite{hecht2012new}).

The computational domain is defined as  $\left\{ (y_{1},y_{3}) | y_{1} \in [-5,5], y_{3}\in [0,1]\right\}$.  This domain is discretized using a triangular mesh with nearly uniform mesh sizes across the entire domain. In a typical simulation, the mesh consists of 17,594 vertices and 34,286 triangles. The simulation employs P1 elements, which correspond to linear functions defined over the triangles.

To simulate the effective equation \eqref{eq:effective equation} for comparison with the simulation of the governing equation \eqref{eq:NS nondimensional 3D diffusion driven flow rotated coordinate}, we employ the Fourier spectral method, as detailed in the work by \cite{ding2022determinism}, which incorporates a third-order implicit–explicit Runge-Kutta scheme proposed in \cite{pareschi2005implicit}. Specifically, we utilize the explicit Runge-Kutta method for integrating the nonlinear terms, while the diffusion term is handled using the implicit Runge-Kutta method. To ensure a meaningful comparison between our simulations and those based on the complete governing equations, we specify our computational domain as $y_{1}\in [-5, 5]$. Additionally, we enforce no-flux boundary conditions at the endpoints of this interval to guarantee the conservation of mass. The Fourier spectral method is particularly effective for periodic domains, but we encounter no-flux boundary conditions in the $y_{1}$-direction. To overcome this problem, we implement an even extension, mirroring functions at $x=5$,  to establish periodic conditions on the extended domain $[-5, 15]$. The Fourier expansion of the even-extended function yields the cosine expansion of the original function. Each cosine function in the expansion has a zero derivative at the endpoints  ($y_{1}=\pm 5$),  ensuring that the no-flux boundary condition is satisfied on the original domain $[-5,5]$. The original domain comprises $2^{10}+1$ grid points, while the extended domain encompasses $2^{11}$ grid points. The typical grid size is  $0.0098$, and the typical time step size is  $9.7561\times 10^{-5}$.

\bibliographystyle{jfm}

\begin{thebibliography}{75}
\expandafter\ifx\csname natexlab\endcsname\relax\def\natexlab#1{#1}\fi
\def\au#1{#1} \def\ed#1{#1} \def\yr#1{#1}\def\at#1{#1}\def\jt#1{\textit{#1}}
  \def\bt#1{#1}\def\bvol#1{\textbf{#1}} \def\vol#1{#1} \def\pg#1{#1}
  \def\publ#1{#1}\def\arxiv#1{#1}\def\org#1{#1}\def\st#1{\textit{#1}}

\bibitem[Abaid {\em et~al.\/}(2004)Abaid, Adalsteinsson, Agyapong \&
  McLaughlin]{abaid2004internal}
{\sc \au{Abaid, Nicole}, \au{Adalsteinsson, David}, \au{Agyapong, Akua} \&
  \au{McLaughlin, Richard~M}} \yr{2004}  \at{An internal splash: Levitation of
  falling spheres in stratified fluids}.  \jt{Physics of Fluids}
  \bvol{16}~(5),  \pg{1567--1580}.

\bibitem[Achdou {\em et~al.\/}(1998)Achdou, Pironneau \&
  Valentin]{achdou1998effective}
{\sc \au{Achdou, Yves}, \au{Pironneau, Olivier} \& \au{Valentin, Frederic}}
  \yr{1998}  \at{Effective boundary conditions for laminar flows over periodic
  rough boundaries}.  \jt{Journal of Computational Physics}  \bvol{147}~(1),
  \pg{187--218}.

\bibitem[Allshouse(2010)]{allshouse2010novel}
{\sc \au{Allshouse, Michael~R}} \yr{2010}  \at{Novel applications of
  diffusion-driven flow}. PhD thesis, Massachusetts Institute of Technology.

\bibitem[Allshouse {\em et~al.\/}(2010)Allshouse, Barad \&
  Peacock]{allshouse2010propulsion}
{\sc \au{Allshouse, Michael~R}, \au{Barad, Michael~F} \& \au{Peacock, Thomas}}
  \yr{2010}  \at{Propulsion generated by diffusion-driven flow}.  \jt{Nature
  Physics}  \bvol{6}~(7),  \pg{516--519}.

\bibitem[Aref {\em et~al.\/}(2017)Aref, Blake, Budi{\v{s}}i{\'c}, Cardoso,
  Cartwright, Clercx, El~Omari, Feudel, Golestanian, Gouillart {\em
  et~al.\/}]{aref2017frontiers}
{\sc \au{Aref, Hassan}, \au{Blake, John~R}, \au{Budi{\v{s}}i{\'c}, Marko},
  \au{Cardoso, Silvana~SS}, \au{Cartwright, Julyan~HE}, \au{Clercx, Herman~JH},
  \au{El~Omari, Kamal}, \au{Feudel, Ulrike}, \au{Golestanian, Ramin},
  \au{Gouillart, Emmanuelle} \& \au{others}} \yr{2017}  \at{Frontiers of
  chaotic advection}.  \jt{Reviews of Modern Physics}  \bvol{89}~(2),
  \pg{025007}.

\bibitem[Aris(1956)]{aris1956dispersion}
{\sc \au{Aris, Rutherford}} \yr{1956}  \at{On the dispersion of a solute in a
  fluid flowing through a tube}.  \jt{Proceedings of the Royal Society of
  London. Series A. Mathematical and Physical Sciences}  \bvol{235}~(1200),
  \pg{67--77}.

\bibitem[Aulbach \& Wanner(1996)]{aulbach1996integral}
{\sc \au{Aulbach, Bernd} \& \au{Wanner, Thomas}} \yr{1996}  \at{Integral
  manifolds for {Carath{\'e}odory} type differential equations in {Banach}
  spaces}.  \jt{Six lectures on dynamical systems}  \bvol{2}.

\bibitem[Aulbach \& Wanner(1999)]{aulbach1999invariant}
{\sc \au{Aulbach, Bernd} \& \au{Wanner, Thomas}} \yr{1999}  \at{Invariant
  foliations for {Carath{\'e}odory} type differential equations in banach
  spaces}.  \jt{Advances of Stability Theory at the End of XX Century’,
  Gordon \& Breach Publishers.} .

\bibitem[Balakotaiah {\em et~al.\/}(1995)Balakotaiah, Chang \&
  Smith]{balakotaiah1995dispersion}
{\sc \au{Balakotaiah, Vemuri}, \au{Chang, Hsueh-Chia} \& \au{Smith, FT}}
  \yr{1995}  \at{Dispersion of chemical solutes in chromatographs and
  reactors}.  \jt{Philosophical Transactions of the Royal Society of London.
  Series A: Physical and Engineering Sciences}  \bvol{351}~(1695),
  \pg{39--75}.

\bibitem[Braunsfurth {\em et~al.\/}(1997)Braunsfurth, Skeldon, Juel, Mullin \&
  Riley]{braunsfurth1997free}
{\sc \au{Braunsfurth, MG}, \au{Skeldon, AC}, \au{Juel, A}, \au{Mullin, T} \&
  \au{Riley, DS}} \yr{1997}  \at{Free convection in liquid gallium}.
  \jt{Journal of Fluid Mechanics}  \bvol{342},  \pg{295--314}.

\bibitem[Camassa {\em et~al.\/}(2022)Camassa, Ding, McLaughlin, Overman, Parker
  \& Vaidya]{camassa2022critical}
{\sc \au{Camassa, Roberto}, \au{Ding, Lingyun}, \au{McLaughlin, Richard~M},
  \au{Overman, Robert}, \au{Parker, Richard} \& \au{Vaidya, Ashwin}} \yr{2022}
  \at{Critical density triplets for the arrestment of a sphere falling in a
  sharply stratified fluid}.  \jt{Recent Advances in Mechanics and
  Fluid-Structure Interaction with Applications: The Bong Jae Chung Memorial
  Volume}  \pg{p.~69}.

\bibitem[Camassa {\em et~al.\/}(2019)Camassa, Harris, Hunt, Kilic \&
  McLaughlin]{camassa2019first}
{\sc \au{Camassa, Roberto}, \au{Harris, Daniel~M}, \au{Hunt, Robert},
  \au{Kilic, Zeliha} \& \au{McLaughlin, Richard~M}} \yr{2019}  \at{A
  first-principle mechanism for particulate aggregation and self-assembly in
  stratified fluids}.  \jt{Nature communications}  \bvol{10}~(1),  \pg{1--8}.

\bibitem[Carney \& Engquist(2022)]{carney2022heterogeneous}
{\sc \au{Carney, Sean~P} \& \au{Engquist, Bj{\"o}rn}} \yr{2022}
  \at{Heterogeneous multiscale methods for rough-wall laminar viscous flow}.
  \jt{Communications in Mathematical Sciences}  \bvol{20}~(8),
  \pg{2069--2106}.

\bibitem[Carr(1979)]{carr1979applications}
{\sc \au{Carr, Jack}} \yr{1979} {\em Applications of centre manifold theory\/}.
   \publ{Lefschetz Center for Dynamical Systems, Division of Applied
  Mathematics}.

\bibitem[Cenedese \& Adduce(2008)]{cenedese2008mixing}
{\sc \au{Cenedese, Claudia} \& \au{Adduce, Claudia}} \yr{2008}  \at{Mixing in a
  density-driven current flowing down a slope in a rotating fluid}.
  \jt{Journal of Fluid Mechanics}  \bvol{604},  \pg{369--388}.

\bibitem[Chang \& Santiago(2023)]{chang2023taylor}
{\sc \au{Chang, Ray} \& \au{Santiago, Juan~G}} \yr{2023}  \at{Taylor dispersion
  in arbitrarily shaped axisymmetric channels}.  \jt{Journal of Fluid
  Mechanics}  \bvol{976},  \pg{A30}.

\bibitem[Chashechkin(2018)]{chashechkin2018waves}
{\sc \au{Chashechkin, Yuli~D}} \yr{2018}  \at{Waves, vortices and ligaments in
  fluid flows of different scales}.  \jt{Physics \& Astronomy International
  Journal}  \bvol{2}~(2),  \pg{105--108}.

\bibitem[Chashechkin \& Mitkin(2004)]{chashechkin2004visual}
{\sc \au{Chashechkin, Yu~D} \& \au{Mitkin, Vladimir~V}} \yr{2004}  \at{A visual
  study on flow pattern around the strip moving uniformly in a continuously
  stratified fluid}.  \jt{Journal of Visualization}  \bvol{7}~(2),
  \pg{127--134}.

\bibitem[Chatwin(1970)]{chatwin1970approach}
{\sc \au{Chatwin, PC}} \yr{1970}  \at{The approach to normality of the
  concentration distribution of a solute in a solvent flowing along a straight
  pipe}.  \jt{Journal of Fluid Mechanics}  \bvol{43}~(2),  \pg{321--352}.

\bibitem[Dell \& Pratt(2015)]{dell2015diffusive}
{\sc \au{Dell, RW} \& \au{Pratt, LJ}} \yr{2015}  \at{Diffusive boundary layers
  over varying topography}.  \jt{Journal of Fluid Mechanics}  \bvol{769},
  \pg{635--653}.

\bibitem[Dimitrieva(2019)]{dimitrieva2019stratified}
{\sc \au{Dimitrieva, NF}} \yr{2019}  \at{Stratified flow structure near the
  horizontal wedge}.  \jt{Fluid Dynamics}  \bvol{54},  \pg{940--947}.

\bibitem[Ding(2022)]{ding2022scalar}
{\sc \au{Ding, Lingyun}} \yr{2022}  \at{Scalar transport and mixing}. PhD
  thesis, The University of North Carolina at Chapel Hill.

\bibitem[Ding(2023)]{ding2023shear}
{\sc \au{Ding, Lingyun}} \yr{2023}  \at{Shear dispersion of multispecies
  electrolyte solutions in the channel domain}.  \jt{Journal of Fluid
  Mechanics}  \bvol{970},  \pg{A27}.

\bibitem[Ding {\em et~al.\/}(2021)Ding, Hunt, McLaughlin \&
  Woodie]{ding2021enhanced}
{\sc \au{Ding, Lingyun}, \au{Hunt, Robert}, \au{McLaughlin, Richard~M} \&
  \au{Woodie, Hunter}} \yr{2021}  \at{Enhanced diffusivity and skewness of a
  diffusing tracer in the presence of an oscillating wall}.  \jt{Research in
  the Mathematical Sciences}  \bvol{8}~(3),  \pg{1--29}.

\bibitem[Ding \& McLaughlin(2022{\natexlab{{\em a\/}}})]{ding2022determinism}
{\sc \au{Ding, Lingyun} \& \au{McLaughlin, Richard~M}} \yr{2022{\natexlab{{\em
  a\/}}}}  \at{Determinism and invariant measures for diffusing passive scalars
  advected by unsteady random shear flows}.  \jt{Physical Review Fluids}
  \bvol{7}~(7),  \pg{074502}.

\bibitem[Ding \& McLaughlin(2022{\natexlab{{\em b\/}}})]{ding2022ergodicity}
{\sc \au{Ding, Lingyun} \& \au{McLaughlin, Richard~M}} \yr{2022{\natexlab{{\em
  b\/}}}}  \at{Ergodicity and invariant measures for a diffusing passive scalar
  advected by a random channel shear flow and the connection between the
  {Kraichnan}-{Majda} model and {Taylor}-{Aris} dispersion}.  \jt{Physica D:
  Nonlinear Phenomena}  \bvol{432},  \pg{133118}.

\bibitem[Ding \& McLaughlin(2023)]{ding2023dispersion}
{\sc \au{Ding, Lingyun} \& \au{McLaughlin, Richard~M.}} \yr{2023}
  \at{Dispersion induced by unsteady diffusion-driven flow in a parallel-plate
  channel}.  \jt{Phys. Rev. Fluids}  \bvol{8},  \pg{084501}.

\bibitem[Drake {\em et~al.\/}(2020)Drake, Ferrari \& Callies]{drake2020abyssal}
{\sc \au{Drake, Henri~F}, \au{Ferrari, Raffaele} \& \au{Callies, J{\"o}rn}}
  \yr{2020}  \at{Abyssal circulation driven by near-boundary mixing: Water mass
  transformations and interior stratification}.  \jt{Journal of Physical
  Oceanography}  \bvol{50}~(8),  \pg{2203--2226}.

\bibitem[French(2017)]{french2017diffusion}
{\sc \au{French, Alison}} \yr{2017}  \at{Diffusion-driven flow in three
  dimensions}. PhD thesis, Monash University.

\bibitem[Gill(1967)]{gill1967note}
{\sc \au{Gill, WN}} \yr{1967}  \at{A note on the solution of transient
  dispersion problems}.  \jt{Proceedings of the Royal Society of London. Series
  A. Mathematical and Physical Sciences}  \bvol{298}~(1454),  \pg{335--339}.

\bibitem[Grayer {\em et~al.\/}(2020)Grayer, Yalim, Welfert \&
  Lopez]{grayer2020dynamics}
{\sc \au{Grayer, Hezekiah}, \au{Yalim, Jason}, \au{Welfert, Bruno~D} \&
  \au{Lopez, Juan~M}} \yr{2020}  \at{Dynamics in a stably stratified tilted
  square cavity}.  \jt{Journal of Fluid Mechanics}  \bvol{883}.

\bibitem[Grayer~II {\em et~al.\/}(2021)Grayer~II, Yalim, Welfert \&
  Lopez]{grayer2021stably}
{\sc \au{Grayer~II, Hezekiah}, \au{Yalim, Jason}, \au{Welfert, Bruno~D} \&
  \au{Lopez, Juan~M}} \yr{2021}  \at{Stably stratified square cavity subjected
  to horizontal oscillations: responses to small amplitude forcing}.
  \jt{journal of fluid mechanics}  \bvol{915},  \pg{A85}.

\bibitem[Hall(1924)]{hall1924densities}
{\sc \au{Hall, Ralph~E}} \yr{1924}  \at{The densities and specific volumes of
  sodium chloride solutions at 25°}.  \jt{Journal of the Washington Academy of
  Sciences}  \bvol{14}~(8),  \pg{167--173}.

\bibitem[Hecht(2012)]{hecht2012new}
{\sc \au{Hecht, Fr{\'e}d{\'e}ric}} \yr{2012}  \at{New development in
  freefem++}.  \jt{Journal of numerical mathematics}  \bvol{20}~(3-4),
  \pg{251--266}.

\bibitem[Hecht {\em et~al.\/}(2005)Hecht, Pironneau, Le~Hyaric \&
  Ohtsuka]{hecht2005freefem++}
{\sc \au{Hecht, Fr{\'e}d{\'e}ric}, \au{Pironneau, Olivier}, \au{Le~Hyaric, A}
  \& \au{Ohtsuka, K}} \yr{2005}  \at{Freefem++ manual}.  \jt{Laboratoire
  Jacques Louis Lions} .

\bibitem[Heitz {\em et~al.\/}(2005)Heitz, Peacock \&
  Stocker]{heitz2005optimizing}
{\sc \au{Heitz, Renaud}, \au{Peacock, Thomas} \& \au{Stocker, Roman}} \yr{2005}
   \at{Optimizing diffusion-driven flow in a fissure}.  \jt{Physics of Fluids}
  \bvol{17}~(12),  \pg{128104}.

\bibitem[Holmes {\em et~al.\/}(2019)Holmes, de~Lavergne \&
  McDougall]{holmes2019tracer}
{\sc \au{Holmes, Ryan~M}, \au{de~Lavergne, Casimir} \& \au{McDougall,
  Trevor~J}} \yr{2019}  \at{Tracer transport within abyssal mixing layers}.
  \jt{Journal of Physical Oceanography}  \bvol{49}~(10),  \pg{2669--2695}.

\bibitem[Kistovich \& Chashechkin(1993)]{kistovich1993structure}
{\sc \au{Kistovich, AV} \& \au{Chashechkin, Yu~D}} \yr{1993}  \at{The structure
  of transient boundary flow along an inclined plane in a continuously
  stratified medium}.  \jt{Journal of Applied Mathematics and Mechanics}
  \bvol{57}~(4),  \pg{633--639}.

\bibitem[Kondic(2003)]{kondic2003instabilities}
{\sc \au{Kondic, Lou}} \yr{2003}  \at{Instabilities in gravity driven flow of
  thin fluid films}.  \jt{Siam review}  \bvol{45}~(1),  \pg{95--115}.

\bibitem[Levitsky {\em et~al.\/}(2019)Levitsky, Dimitrieva \&
  Chashechkin]{levitsky2019visualization}
{\sc \au{Levitsky, VV}, \au{Dimitrieva, NF} \& \au{Chashechkin, Yu~D}}
  \yr{2019}  \at{Visualization of the self-motion of a free wedge of neutral
  buoyancy in a tank filled with a continuously stratified fluid and
  calculation of perturbations of the fields of physical quantities putting the
  body in motion}.  \jt{Fluid Dynamics}  \bvol{54},  \pg{948--957}.

\bibitem[Lin {\em et~al.\/}(2011)Lin, Thiffeault \& Doering]{lin2011optimal}
{\sc \au{Lin, Zhi}, \au{Thiffeault, Jean-Luc} \& \au{Doering, Charles~R}}
  \yr{2011}  \at{Optimal stirring strategies for passive scalar mixing}.
  \jt{Journal of Fluid Mechanics}  \bvol{675},  \pg{465--476}.

\bibitem[Linden(1979)]{linden1979mixing}
{\sc \au{Linden, PF}} \yr{1979}  \at{Mixing in stratified fluids}.
  \jt{Geophysical \& Astrophysical Fluid Dynamics}  \bvol{13}~(1),  \pg{3--23}.

\bibitem[Linden \& Weber(1977)]{linden1977formation}
{\sc \au{Linden, PF} \& \au{Weber, JE}} \yr{1977}  \at{The formation of layers
  in a double-diffusive system with a sloping boundary}.  \jt{Journal of Fluid
  Mechanics}  \bvol{81}~(4),  \pg{757--773}.

\bibitem[Magnaudet \& Mercier(2020)]{magnaudet2020particles}
{\sc \au{Magnaudet, Jacques} \& \au{Mercier, Matthieu~J}} \yr{2020}
  \at{Particles, drops, and bubbles moving across sharp interfaces and
  stratified layers}.  \jt{Annual Review of Fluid Mechanics}  \bvol{52},
  \pg{61--91}.

\bibitem[Mercer \& Roberts(1990)]{mercer1990centre}
{\sc \au{Mercer, GN} \& \au{Roberts, AJ}} \yr{1990}  \at{A centre manifold
  description of contaminant dispersion in channels with varying flow
  properties}.  \jt{SIAM Journal on Applied Mathematics}  \bvol{50}~(6),
  \pg{1547--1565}.

\bibitem[Mercier {\em et~al.\/}(2014)Mercier, Ardekani, Allshouse, Doyle \&
  Peacock]{mercier2014self}
{\sc \au{Mercier, Matthieu~J}, \au{Ardekani, Arezoo~M}, \au{Allshouse,
  Michael~R}, \au{Doyle, Brian} \& \au{Peacock, Thomas}} \yr{2014}
  \at{Self-propulsion of immersed objects via natural convection}.
  \jt{Physical review letters}  \bvol{112}~(20),  \pg{204501}.

\bibitem[More \& Ardekani(2023)]{more2023motion}
{\sc \au{More, Rishabh~V} \& \au{Ardekani, Arezoo~M}} \yr{2023}  \at{Motion in
  stratified fluids}.  \jt{Annual Review of Fluid Mechanics}  \bvol{55},
  \pg{157--192}.

\bibitem[Newman(1984)]{newman1984lyapunov}
{\sc \au{Newman, William~I}} \yr{1984}  \at{A lyapunov functional for the
  evolution of solutions to the porous medium equation to self-similarity. i}.
  \jt{Journal of Mathematical Physics}  \bvol{25}~(10),  \pg{3120--3123}.

\bibitem[Oerlemans \& Grisogono(2002)]{oerlemans2002glacier}
{\sc \au{Oerlemans, J} \& \au{Grisogono, B}} \yr{2002}  \at{Glacier winds and
  parameterisation of the related surface heat fluxes}.  \jt{Tellus A: Dynamic
  Meteorology and Oceanography}  \bvol{54}~(5),  \pg{440--452}.

\bibitem[Page(2011{\natexlab{{\em a\/}}})]{page2011combined}
{\sc \au{Page, Michael~A}} \yr{2011{\natexlab{{\em a\/}}}}  \at{Combined
  diffusion-driven and convective flow in a tilted square container}.
  \jt{Physics of Fluids}  \bvol{23}~(5),  \pg{056602}.

\bibitem[Page(2011{\natexlab{{\em b\/}}})]{page2011steady}
{\sc \au{Page, Michael~A}} \yr{2011{\natexlab{{\em b\/}}}}  \at{Steady
  diffusion-driven flow in a tilted square container}.  \jt{The Quarterly
  Journal of Mechanics \& Applied Mathematics}  \bvol{64}~(3),  \pg{319--348}.

\bibitem[Pareschi \& Russo(2005)]{pareschi2005implicit}
{\sc \au{Pareschi, Lorenzo} \& \au{Russo, Giovanni}} \yr{2005}
  \at{Implicit--explicit {Runge--Kutta} schemes and applications to hyperbolic
  systems with relaxation}.  \jt{Journal of Scientific computing}
  \bvol{25}~(1),  \pg{129--155}.

\bibitem[Pavliotis \& Stuart(2008)]{pavliotis2008multiscale}
{\sc \au{Pavliotis, Grigoris} \& \au{Stuart, Andrew}} \yr{2008} {\em Multiscale
  methods: averaging and homogenization\/}.  \publ{Springer Science \& Business
  Media}.

\bibitem[Phillips(1970)]{phillips1970flows}
{\sc \au{Phillips, OM}} \yr{1970} On flows induced by diffusion in a stably
  stratified fluid.  \bt{In {\em Deep Sea Research and Oceanographic
  Abstracts\/}}, ,  \vol{vol.~17},  \pg{pp. 435--443}. Elsevier.

\bibitem[Polyanin \& Zaitsev(2012)]{polyanin2012handbook}
{\sc \au{Polyanin, AD} \& \au{Zaitsev, VF}} \yr{2012}  \at{Handbook of
  nonlinear partial differential equations} .

\bibitem[Prandtl {\em et~al.\/}(1942)Prandtl, Oswatitsch \&
  Wieghardt]{prandtl1942fuhrer}
{\sc \au{Prandtl, L}, \au{Oswatitsch, K} \& \au{Wieghardt, K}} \yr{1942}
  \at{F{\"u}hrer durch die str{\"o}mungslehre (essentials of fluid mechanics)}.
   \jt{Fried. Vieweg \& Sohn}  \pg{pp. 105--108}.

\bibitem[Roberts(1988)]{roberts1988application}
{\sc \au{Roberts, AJ}} \yr{1988}  \at{The application of centre-manifold theory
  to the evolution of system which vary slowly in space}.  \jt{The ANZIAM
  Journal}  \bvol{29}~(4),  \pg{480--500}.

\bibitem[Roberts(1996)]{roberts1996low}
{\sc \au{Roberts, AJ}} \yr{1996}  \at{Low-dimensional models of thin film fluid
  dynamics}.  \jt{Physics Letters A}  \bvol{212}~(1-2),  \pg{63--71}.

\bibitem[Roberts(2015)]{roberts2015macroscale}
{\sc \au{Roberts, AJ}} \yr{2015}  \at{Macroscale, slowly varying, models emerge
  from the microscale dynamics}.  \jt{IMA Journal of Applied Mathematics}
  \bvol{80}~(5),  \pg{1492--1518}.

\bibitem[Roberts \& Li(2006)]{roberts2006accurate}
{\sc \au{Roberts, AJ} \& \au{Li, Zhenquan}} \yr{2006}  \at{An accurate and
  comprehensive model of thin fluid flows with inertia on curved substrates}.
  \jt{Journal of Fluid Mechanics}  \bvol{553},  \pg{33--73}.

\bibitem[Roberts(2014)]{roberts2014model}
{\sc \au{Roberts, Anthony~John}} \yr{2014} {\em Model emergent dynamics in
  complex systems\/}, ,  \vol{vol.~20}.  \publ{SIAM}.

\bibitem[Roggeveen {\em et~al.\/}(2023)Roggeveen, Stone \&
  Kurzthaler]{roggeveen2023transport}
{\sc \au{Roggeveen, James~V}, \au{Stone, Howard~A} \& \au{Kurzthaler,
  Christina}} \yr{2023}  \at{Transport of a passive scalar in wide channels
  with surface topography: An asymptotic theory}.  \jt{Journal of Physics:
  Condensed Matter}  \bvol{35}~(27),  \pg{274003}.

\bibitem[S{\'a}nchez {\em et~al.\/}(2005)S{\'a}nchez, Higuera \&
  Medina]{sanchez2005natural}
{\sc \au{S{\'a}nchez, F}, \au{Higuera, FJ} \& \au{Medina, A}} \yr{2005}
  \at{Natural convection in tilted cylindrical cavities embedded in rocks}.
  \jt{Physical Review E}  \bvol{71}~(6),  \pg{066308}.

\bibitem[Shaughnessy \& Van~Gilder(1995)]{shaughnessy1995low}
{\sc \au{Shaughnessy, Edward~J} \& \au{Van~Gilder, James~W}} \yr{1995}  \at{Low
  rayleigh number conjugate convection in straight inclined fractures in rock}.
   \jt{Numerical Heat Transfer, Part A: Applications}  \bvol{28}~(4),
  \pg{389--408}.

\bibitem[Smith(1982)]{smith1982contaminant}
{\sc \au{Smith, Ronald}} \yr{1982}  \at{Contaminant dispersion in oscillatory
  flows}.  \jt{Journal of Fluid Mechanics}  \bvol{114},  \pg{379--398}.

\bibitem[Taylor(1953)]{taylor1953dispersion}
{\sc \au{Taylor, Geoffrey~Ingram}} \yr{1953}  \at{Dispersion of soluble matter
  in solvent flowing slowly through a tube}.  \jt{Proceedings of the Royal
  Society of London. Series A. Mathematical and Physical Sciences}
  \bvol{219}~(1137),  \pg{186--203}.

\bibitem[Thiffeault(2012)]{thiffeault2012using}
{\sc \au{Thiffeault, Jean-Luc}} \yr{2012}  \at{Using multiscale norms to
  quantify mixing and transport}.  \jt{Nonlinearity}  \bvol{25}~(2),  \pg{R1}.

\bibitem[Thomas \& Camassa(2023)]{thomas2023self}
{\sc \au{Thomas, Jim} \& \au{Camassa, Roberto}} \yr{2023}  \at{Self-induced
  flow over a cylinder in a stratified fluid}.  \jt{Journal of Fluid Mechanics}
   \bvol{964},  \pg{A38}.

\bibitem[Van~Dyke(1987)]{van1987slow}
{\sc \au{Van~Dyke, Milton}} \yr{1987}  \at{Slow variations in continuum
  mechanics}.  \jt{Advances in applied mechanics}  \bvol{25},  \pg{1--45}.

\bibitem[Vitagliano \& Lyons(1956)]{vitagliano1956diffusion}
{\sc \au{Vitagliano, V} \& \au{Lyons, Phillip~A}} \yr{1956}  \at{Diffusion
  coefficients for aqueous solutions of sodium chloride and barium chloride}.
  \jt{Journal of the American Chemical Society}  \bvol{78}~(8),
  \pg{1549--1552}.

\bibitem[Woods \& Linz(1992)]{woods1992natural}
{\sc \au{Woods, Andrew~W} \& \au{Linz, Stefan~J}} \yr{1992}  \at{Natural
  convection and dispersion in a tilted fracture}.  \jt{Journal of Fluid
  Mechanics}  \bvol{241},  \pg{59--74}.

\bibitem[Wu \& Chen(2014)]{wu2014approach}
{\sc \au{Wu, Zi} \& \au{Chen, GQ}} \yr{2014}  \at{Approach to transverse
  uniformity of concentration distribution of a solute in a solvent flowing
  along a straight pipe}.  \jt{Journal of Fluid Mechanics}  \bvol{740},
  \pg{196--213}.

\bibitem[Wunsch(1970)]{wunsch1970oceanic}
{\sc \au{Wunsch, Carl}} \yr{1970} On oceanic boundary mixing.  \bt{In {\em Deep
  Sea Research and Oceanographic Abstracts\/}}, ,  \vol{vol.~17},  \pg{pp.
  293--301}. Elsevier.

\bibitem[Young \& Jones(1991)]{young1991shear}
{\sc \au{Young, WR~a} \& \au{Jones, Scott}} \yr{1991}  \at{Shear dispersion}.
  \jt{Physics of Fluids A: Fluid Dynamics}  \bvol{3}~(5),  \pg{1087--1101}.

\bibitem[Zagumennyi \& Dimitrieva(2016)]{zagumennyi2016diffusion}
{\sc \au{Zagumennyi, Ia~V} \& \au{Dimitrieva, NF}} \yr{2016}  \at{Diffusion
  induced flow on a wedge-shaped obstacle}.  \jt{Physica Scripta}
  \bvol{91}~(8),  \pg{084002}.

\end{thebibliography}

\end{document}